\documentclass[useAMS,usenatbib]{mn2e}
\usepackage{graphicx}
\usepackage{txfonts}
\usepackage{natbib}

\title[Search for pulsars and fast transients in M31] {A search for
  radio pulsars and fast transients in M31 using the WSRT}

\author[E. Rubio-Herrera, B.~W. Stappers, J.~W.~T. Hessels \& R. Braun]{E.~Rubio--Herrera \thanks{e--mail: erubio@astro.unam.mx}$^{1,2}$, B.~W.~Stappers$^3$, J.~W.~T.~Hessels$^{4,2}$, R. Braun$^{5}$  \\
  $^1$Instituto de Astronom\'{\i}a Universidad Nacional Aut\'onoma de
  M\'exico, Apartado Postal 70--264 CP 04510 M\'exico DF M\'exico. \\
  $^2$Astronomical Institute ``Anton Pannekoek'', University of
  Amsterdam, Science Park 904, 1098 XH Amsterdam, The Netherlands.\\
  $^3$Jodrell Bank Centre for Astrophysics, School of Physics and
  Astronomy, University of Manchester, Manchester M13 9PL, United
  Kingdom. \\
  $^4$ASTRON, the Netherlands Institute for Radio Astronomy, Postbus 2,
  7990 AA, Dwingeloo, The Netherlands. \\
  $^5$Australia Telescope National Facility P.O. Box 76 AU Epping NSW
  1710 Australia.}
  
\date{Released 2012 xxx XX}

\pagerange{\pageref{firstpage}--\pageref{lastpage}} \pubyear{2012}

\def\LaTeX{L\kern-.36em\raise.3ex\hbox{a}\kern-.15em
    T\kern-.1667em\lower.7ex\hbox{E}\kern-.125emX}

\begin{document}
\label{firstpage}
\maketitle
\begin{abstract}

We present the results of the most sensitive and comprehensive
  survey yet undertaken for radio pulsars and fast transients in the
Andromeda galaxy (M31) and its satellites, using the Westerbork
Synthesis Radio Telescope (WSRT) at a central frequency of
328 MHz. We used the WSRT in a special configuration called
{\tt 8gr8} (eight--grate) mode, which provides a large instantaneous
field-of-view, about 5 square degrees per pointing, with good
sensitivity, long dwell times (up to 8 hours per pointing), and good
spatial resolution (a few arc minutes) for locating sources.  We
have searched for both periodicities and single pulses in our data,
aiming to detect bright, persistent radio pulsars and rotating radio
transients (RRATs) of either Galactic or extragalactic origin. Our
searches did not reveal any confirmed periodic signals or bright
single bursts from (potentially) cosmological distances. However, we
do report the detection of several single pulse events, some repeating
at the same dispersion measure, which could potentially originate from
neutron stars in M31. One in particular was seen multiple times,
including a burst of six pulses in 2000 seconds, at a dispersion
measure of 54.7 pc cm$^{-3}$, which potentially places the origin
  of this source outside of our Galaxy.  Our results are
compared to a range of hypothetical populations of pulsars and RRATs
in M31 and allow us to constrain the luminosity function of pulsars in
M31. They also show that, unless the pulsar population in  M31 is much 
dimmer than in our Galaxy, there is no need to invoke any violation  
of the inverse square law of the distance for pulsar fluxes.

\end{abstract}


\section{Introduction}

Extragalactic radio pulsars, i.e. pulsars located in galaxies beyond
our own Galaxy and its satellites, have yet to be detected.  This is
primarily because they are at least 100 times more distant than the
majority of the known population of radio pulsars in our Galaxy. To
date, the most distant pulsars detected have been found inside the
Large Magellanic Cloud (LMC) and the Small Magellanic Cloud (SMC),
located at distances of 49 and 57 kpc respectively
(e.g. \citealt{mfl+06}). Several attempts to detect extragalactic
radio pulsars have been made in the past.  \cite{le+80} reported the
detection of highly dispersed radio pulses from M87 with a duration of
50 ms.  These findings however were not confirmed by \cite{hcd+81},
\cite{meg+81} or \cite{tbd+81}.

The brightest radio bursts known are the giant pulses seen from the
Crab pulsar (\citealt{cbh+02}) and a handful of other neutron
stars. These pulses are very narrow (in some cases with structure
still unresolved at 2-ns time resolution) and very bright ($>10^5$ Jy)
as reported by \cite{hkwe03}.  Due to their brightness, giant pulses
may in principle be detected in other galaxies and \cite{mc+06}
performed searches for giant pulses from M33, the LMC, NGC253, NGC300,
Fornax, NGC6300 and NGC7793, but with no confirmed detections.  More
recently, \cite{bbc+04} undertook a search for pulsars in M33 at
1400 MHz, also with null results.

Generally, pulsar surveys use two complementary search techniques:
Fourier-based periodicity searches as well as searches for dispersed single
pulses.  Periodicity searches are poorly optimized for sources that
exhibit pulses with a wide range of intensities, as comprehensively
discussed by \cite{cm+06}. Single pulse searches can be more powerful
than periodicity searches if the source emits bright but rare pulses
and/or shows a shallow power-law pulse energy distribution. In these
searches, individual bright, dispersed pulses are recorded and
subsequently analyzed to establish a possible underlying periodicity,
which does not necessarily show up in a periodicity search.

In recent years, single pulse search methods have become standard
practice and are now employed in most of the on-going pulsar surveys.
This has resulted in two important discoveries. The first is the
population of rotating radio transients (RRATs, \citealt{mll+06}),
objects which are characterized by their sporadic, bright radio
pulses, and which possibly represent a subset of the total radio
pulsar population.  \cite{mll+06} show that RRATs are among the most
luminous radio sources known when they emit ($L_{1400}\sim$ 1--3.6 Jy
kpc$^{2}$), though they are still weaker than the giant pulses seen
from, e.g., the Crab pulsar and PSR B1937+21.  RRATs can account for a
considerable fraction of the active radio-emitting neutron stars in
our Galaxy and their high luminosity may make it possible to detect
them in nearby galaxies. The second discovery is the detection of
what appears to be a single radio pulse of
extragalactic origin: the ``Lorimer Burst'' (\citealt{lbm+07}), an
extremely bright ($S_{1400}\sim30$ Jy), $\sim 5$-ms-wide pulse from a
high Galactic latitude and with a high dispersion measure (hereafter
DM) of about $375$ pc cm$^{-3}$, both which suggest an origin well
beyond our galaxy ($d\geq$ 1 Gpc). Such a burst could, e.g., be
associated with the coalescence of two neutron stars
(\citealt{hanly01}) or the evaporation of a black hole
(\citealt{rees77}).  Other detections of this kind (\citealt{kkl+11}) with other
observatories and, especially, where the source position can be better
localized, would help to elucidate the true nature of this phenomenon.
 
Searching for extragalactic pulsars is motivated by gaining a better
understanding of galactic pulsar populations, especially when the star
formation rate and stellar evolution history differ from those in
our Galaxy.  The sensitivity of current radio telescopes is such that
we are in a position to detect only the most luminous pulsars in the
nearby galaxies of our Local Group.  The direct comparison of a pulsar
location with that of the known supernova remnants in M31 such as
those found by ~\cite{glg}, would be difficult due to the small angular
size of these remnants. However, if the pulsar age is also determined,
associations may be possible if the supernova remnants are still
bright (\citealt{nasc88}).  This correlation can help determine
the formation ratio of rotation-powered pulsars versus other
manifestations of neutron stars like magnetars, and other compact
objects like black holes in nearby galaxies.  By measuring the DM of
sources in M31 and comparing them with the models of free electrons
for our Galaxy, we can estimate limits on the contributions to the DM
due to the interstellar medium in M31, assuming that we know the
contribution due to our own galaxy, within the limits of accuracy of
the NE2001 model of \cite{cl02} (the contribution of the intergalactic
medium is likely to be negligible). Finally, by measuring the rotation
measure (RM) it might also be possible to obtain an estimate of the magnetic
field along the line-of-sight to Andromeda.

The detection of pulsars in other Local Group galaxies is challenging,
however. In the case of M31, the disk of the galaxy covers a
significant area of sky ($190^{\prime} \times 60^{\prime}$) and the
distance, $\sim 778$ kpc (\citealt{kkv+3}), means that any pulsar
signal is likely to be extremely faint.  An effective search also
requires long dwell times in order to catch rare, bright bursts.  A
deep search of M31 can be achieved by using the WSRT in a special mode
called {\tt 8gr8}, which makes optimal use of the array's linear
configuration, in combination with the Pulsar Machine (PuMa;
\citealt{vkh+02}) backend.  The {\tt 8gr8} mode provides a
$\sim$2.5-deg-wide field-of-view (FoV), that of the primary beam of the
individual 25-m WSRT dishes, as well as the full sensitivity and
angular resolution of all the dishes combined in a tied-array.  This
enables the possibility of locating new sources with good accuracy (a
few arcminutes) if the source emits bright periodic pulses or, at
least, multiple bright bursts at different hour angles.

In this paper, we present a search of the Andromeda galaxy (M31)
and its satellites (M32, M110 and a few other dwarf galaxies).  In \S
2 we describe the observations and the {\tt 8gr8} technique; in \S 3 we
present the data analysis and describe the search methodology; in \S 4
we present the search results; and in \S 5 we compare our results to
our current understanding of the luminosity distributions of pulsars
and RRATs in our own Galaxy.  Lastly, our conclusions are presented in
\S 6. \\

\begin{figure}
\centering
\hspace*{-0.5cm}
\includegraphics[width=0.5\textwidth, angle=0]{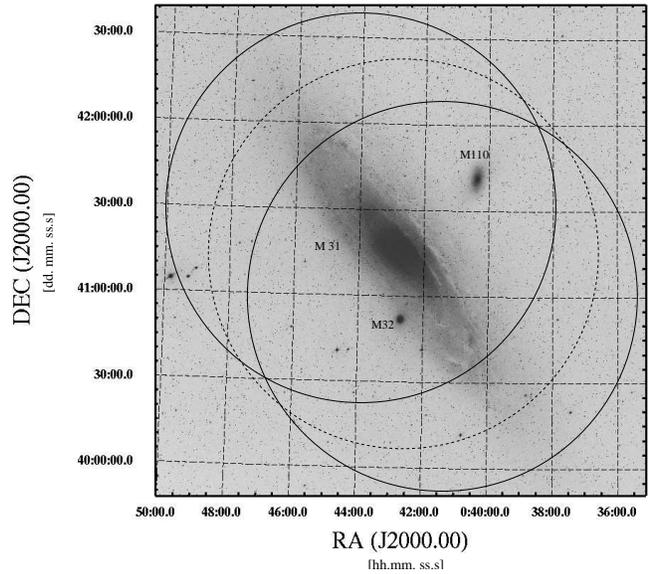}

\caption{A Sloan Digital Sky Survey image of M31.  The circles
  represent the half power beam width of the individual WSRT dishes
  at a frequency of 328 MHz, which cover approximately 5 square
  degrees each. The solid lines represent the two main pointings of
  this survey, PNT01 (south) and PNT02 (north), while the dotted line
  represents PNT03, a complementary pointing centered on the core of
  M31 ~(see Table \ref{table-m31-beams} for reference).  M32 and M110
  also fit within the beams (see labels on the
  figure). \label{m31-beams}}

\end{figure}

\begin{table}
\caption{WSRT observations of M31. $N_{\rm P}$ is the number of pointings at each position.  The start times of the individual pointing are given in Table 7 of Appendix A. \label{table-m31-beams}}
\begin{tabular}{c c c c c}
\hline
\hline
Pointing             & R.A.(J2000.0)  &  DEC.(J2000.0)   & $N_{\rm P}$ & Dwell   \\
Name                 & (h.m.s.)       &  (d.m.s.)        &       & Time (hr)           \\
\hline
M31 PNT01            & 00 41 30       & +41 00 00        &  4    & 8        \\ 
M31 PNT02            & 00 44 00       & +41 30 00        &  4    & 8        \\
M31 PNT03            & 00 42 45       & +41 15 00        &  2    & 8        \\
\hline
\hline
\end{tabular}
\end{table}


\section{WSRT {\tt 8gr8} Observations}

We observed two main pointings along the disk of M31 (solid circles of
Fig.~\ref{m31-beams}) and later also one complementary pointing
centered on the core of the galaxy (dashed circle of
Fig.~\ref{m31-beams}).  Each of the main pointings was observed for 32
h total, divided in to four 8-h observations, separated by intervals
of approximately one day. The complementary pointing was observed for
16 h in two 8-h sessions at a later epoch.  We recorded two
polarizations (summed in quadrature), each with 10 MHz of bandwidth at
a center frequency of 328 MHz. The digital filter-bank PuMa
(\citealt{vkh+02}) was used to write spectra with 256 frequency
channels and $t_{\rm s}=$ 409.6 $\mu$s sampling, producing $2^{26}$
samples per observation. Table 1 summarizes these observations, which
were performed in 2005 October using the WSRT in {\tt 8gr8} mode (see
also \citealt{jsb+05}). In this mode, 12 dishes of 25 m in diameter,
and equally spaced by 144 m, are used to form an array with the
equivalent collecting area of a single 74-m dish.

\begin{figure}
\centering
\hspace*{-0.2cm}
\includegraphics[width=0.3375\textwidth,height=0.45\textwidth,angle=-90]{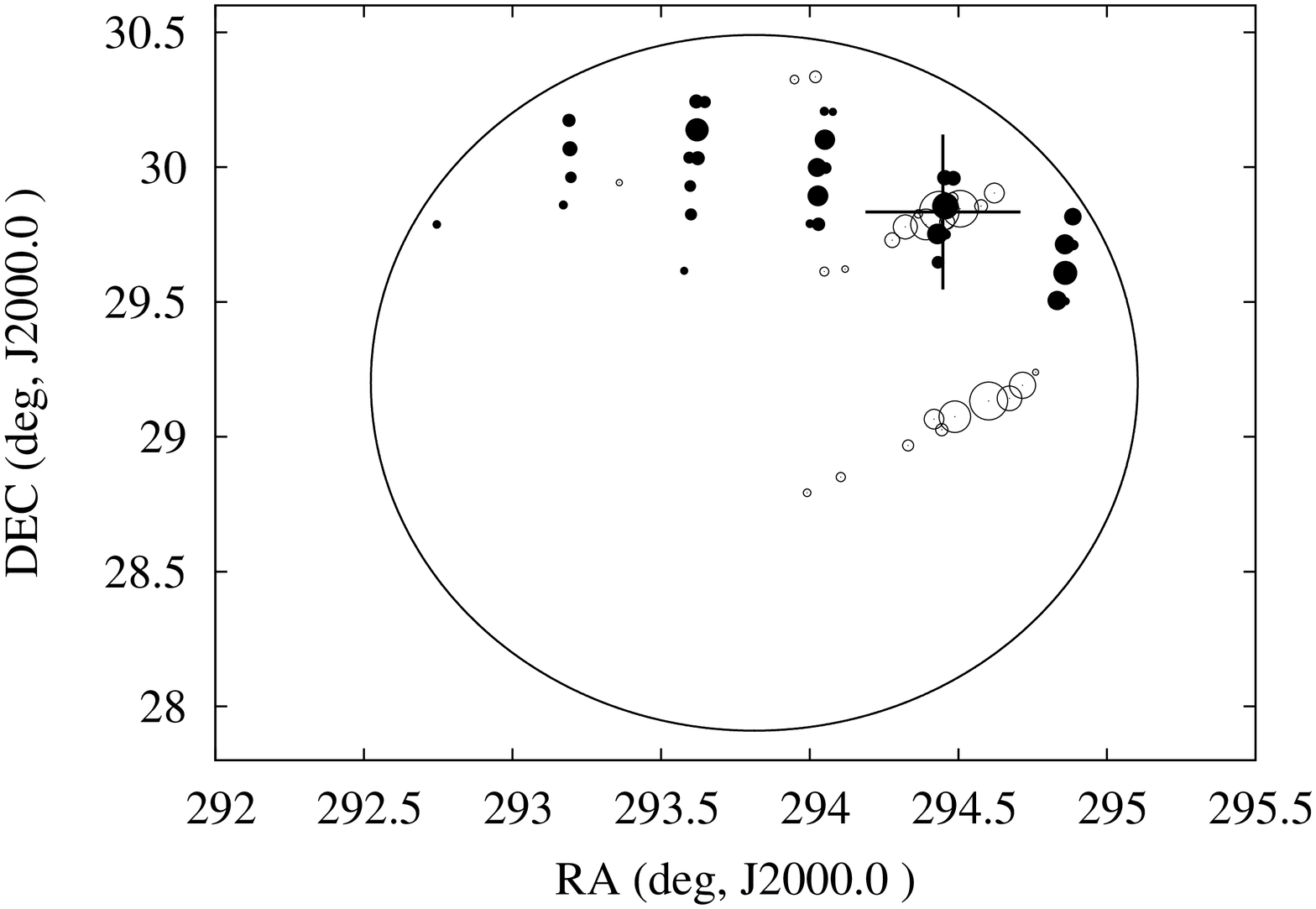}
\vspace*{2mm}
\includegraphics[width=0.355\textwidth,height=0.45\textwidth,angle=-90]{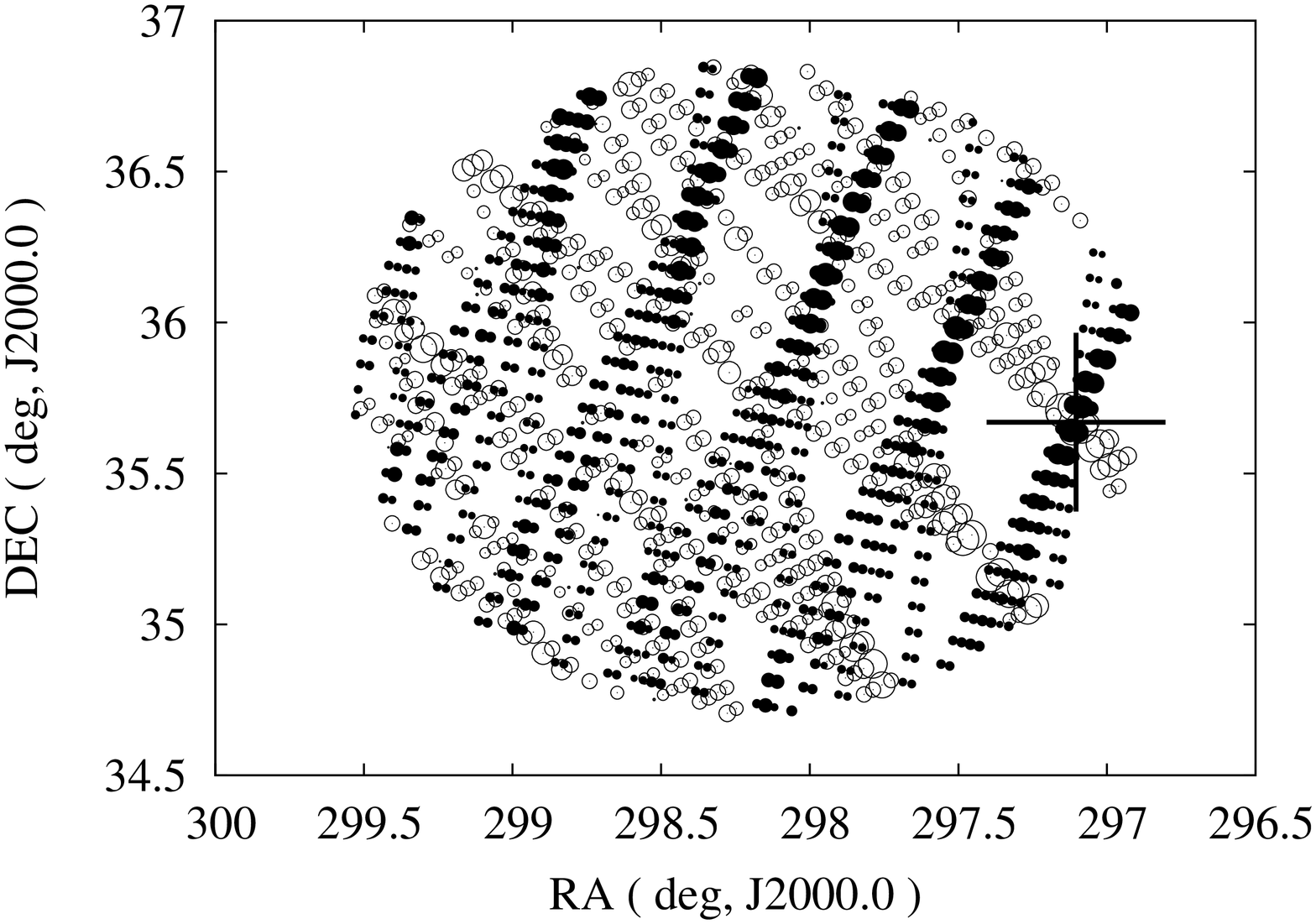}

\caption{An example of how sources can be localized using the
  {\tt 8gr8} method. The panels show two examples of beam plots. The
  upper panel shows the original and confirmation observations in
  which PSR J1937+2950 was detected using a periodicity search
  (position shown with a cross, figure taken from
  \citealt{jsb+05}). These observations were each 2 h long and were
  done at hour angles of HA$_1=138.81^{\circ}$ and
  HA$_2=201.31^{\circ}$. In the lower plot we show the original and
  confirmation observations in which single pulses from the known
  pulsar PSR B1946+35 were detected (position shown with large
  cross). These observations were done at hour angles
  HA$_1=174.45^{\circ}$ and HA$_2=244.45^{\circ}$ respectively.  In
  both panels solid dots represent sub-beams in which the pulsar was
  detected in the first observation while the open circles represent
  detections made in the confirmation observation. The sizes of all
  the circles are proportional to the S/N of the detections. In both
  cases, the two observations at different hour angles allow us to
  accurately locate the source. \label{example-beam-plot}}

\end{figure}

When the signals of the 12 dishes are coherently added, having
corrected for instrumental and geometrical phase delays, we obtain an
elliptically shaped fan beam.  The major axis of the fan beam spans
the entire primary beam of the 25-m dishes, and the minor axis is
inversely proportional to the separation of the furthest dishes. Due
to the regular spacing of the dishes along the East--West axis, they
produce a grating response on the sky with parallel fan beams equally
spaced across the primary beam and separated by $c/(B\nu)$ radians,
where $B$ is the projected baseline between the dishes, $\nu$ is the
observing frequency and $c$ is the speed of light. Each such
collection of fan beams are referred to as a ``grating group'' from now
on. Since the WSRT has eight independent signal chains, which can each
have their own separate phase-tracking center, we can simultaneously
form and record eight grating groups on the sky.

These grating groups allow us to almost fully tessellate the entire
primary beam. At a later stage, the eight grating groups can be
linearly combined in software into $N_{\rm total}$ elliptical sub-beams to
cover the entire FoV of the primary beam by making use of the fact
that the grating groups rotate relative to the sky during the
observation. Each of the sub-beams covers an angular size of
approximately $\sim 1.8$ arcmin, at an observing frequency of
328 MHz, and is spaced in such a way so as to overlap with adjacent
sub-beams at the half power point. The total number of sub-beams
$N_{\rm total}$ varies per observation between 900 and 1100, and depends
on the position of the main beam on the sky (hour angle) and the
length of the observation. Each sub-beam has a comparable sensitivity
to a coherent combination of all 12 dishes. Each of these sub-beams is
searched for dispersed, pulsed signals and single bursts. The
sub-beams allow us to localize sources with modest precision when we
have detections from observations made at different hour angles, since
the different orientations of the fan beams allow one to localize the
source at the crossing points of these beams.  Having a large number
of sub-beams also permits us to distinguish spurious detections from
real ones because genuine single bursts will be detected in multiple
sub-beams.

In the case of periodicity searches, a candidate should appear in
multiple detection regions in an observation, as shown in the upper
panel of Fig. \ref{example-beam-plot}.  The number of beams in which
the candidate is detected depends on the intensity of the source. For
single pulse searches, localization is more complicated because the
source will appear in more sub-beams. A genuine astrophysical
detection should appear in a number of sub-beams $N_{\rm b}$ such that
$N_{\rm max} \geq N_{\rm b} \geq N_{\rm min}$, where $N_{\rm max} \sim 1/8 N_{\rm total}$
and $N_{\rm min} \sim 1\% N_{\rm total}$ with the sub-beams making a specific
pattern within the primary beam as shown in the lower panel of Fig.
\ref{example-beam-plot}. Here we show the detection of 8 pulses in
total, divided into two groups of 4 pulses, each group from a
different observation performed at a significantly different hour
angle.  A detection appearing in too many sub-beams or too many fan
beams ($N_{\rm b} > N_{\rm max}$) is likely to be impulsive radio frequency interference (RFI).
In contrast, a detection that appears in too few sub-beams ($N_{\rm b} < N_{\rm min}$) is
likely to be a statistical fluctuation.

\section{Data reduction}

As we describe in detail below, we searched for both periodic and
sporadic signals. We first discuss the common elements of the
analysis.

The NE2001 model (\citealt{cl02}) predicts\footnote{The systematic
error on this predicted maximum DM is ill-defined, but from other
  examples it is conceivable that this value could be off by 20\% or
  more.} a $\mbox{DM}=55 \pm 10$ pc cm$^{-3}$ for a pulsar at the
outer edge of our Galaxy, along the line-of-sight to M31, which lies
well out of the Galactic Plane ($b=-21^{\circ}$).  To detect sources
located well inside M31, we searched a broad range of trial DMs from
0 to 350 pc cm$^{-3}$. This DM range covers sources located within our
Galaxy ($\mbox{DM}< 45$ pc cm$^{-3}$), as well as objects that might
be located in M31. Our upper limit for the DM trials was set due to
the fact that at this high DM the radio pulses are likely severely
scattered at 328 MHz.  We consider that any value of DM larger than
$\sim$ 45 pc cm$^{-3}$ {\it may} indicate a source location in
M31; assuming that the electron distribution in M31 is not too
  dissimilar to the one in our Galaxy, this DM range would allow us to
  detect sources reasonably deep into M31, even though it is inclined at only about
  $13^{\circ}$ to our line-of-sight (\citealt{sp+04}).  For example, an excess in DM of
  10 pc cm$^{-3}$ (above that contributed by our Galaxy) would imply a
  location about 1 kpc inside the outer edge of Andromeda.

Due to the large data volume, each 8-h observation was sub-divided
into chunks of 1 h for further processing. For each 1-h chunk, we
prepared partially dedispersed time series for each of the 8 grating
groups.  This was done using a modified version of the tree
dedispersion algorithm developed by \cite{tay74}, which divides the
bandwidth of each fan beam into smaller frequency sub-bands where we
assume the dispersive delay to be linear.  From these, we form time
series for all trial DMs (see below), each consisting of $2^{23}$
samples. Lastly, a geometrical correction was applied before combining
the data sets from each of the 8 fan beams into a collection of
sub-beams spanning the whole hour.

Each of these sub-beams was searched for periodicities and bright
pulses. We searched 549 trial DMs, with step sizes of 0.348, 0.697 and
1.392 pc cm$^{-3}$ respectively, each corresponding to the spacing
below the first and second ``diagonal DM''\footnote{At the diagonal
  DM, the dispersion delay across a frequency channel is equal to
  twice the sampling time.}, located at $\mbox{DM}=83$ and 249 pc
cm$^{-3}$ respectively.  The DM step-sizes were chosen so that the
maximum error in the delay between the highest and lowest frequency
channels was equal to the sampling time.  This search used the
{\sc{PSE}} code from the Parkes High-Latitude Survey
(\citealt{ebvb01}), with single-pulse search extensions developed by
us (see \S 3.2).

\begin{figure*}
\begin{center}
\hspace{-0.80cm}
\includegraphics[width=1.05\textwidth,height=0.75\textwidth, angle=0]{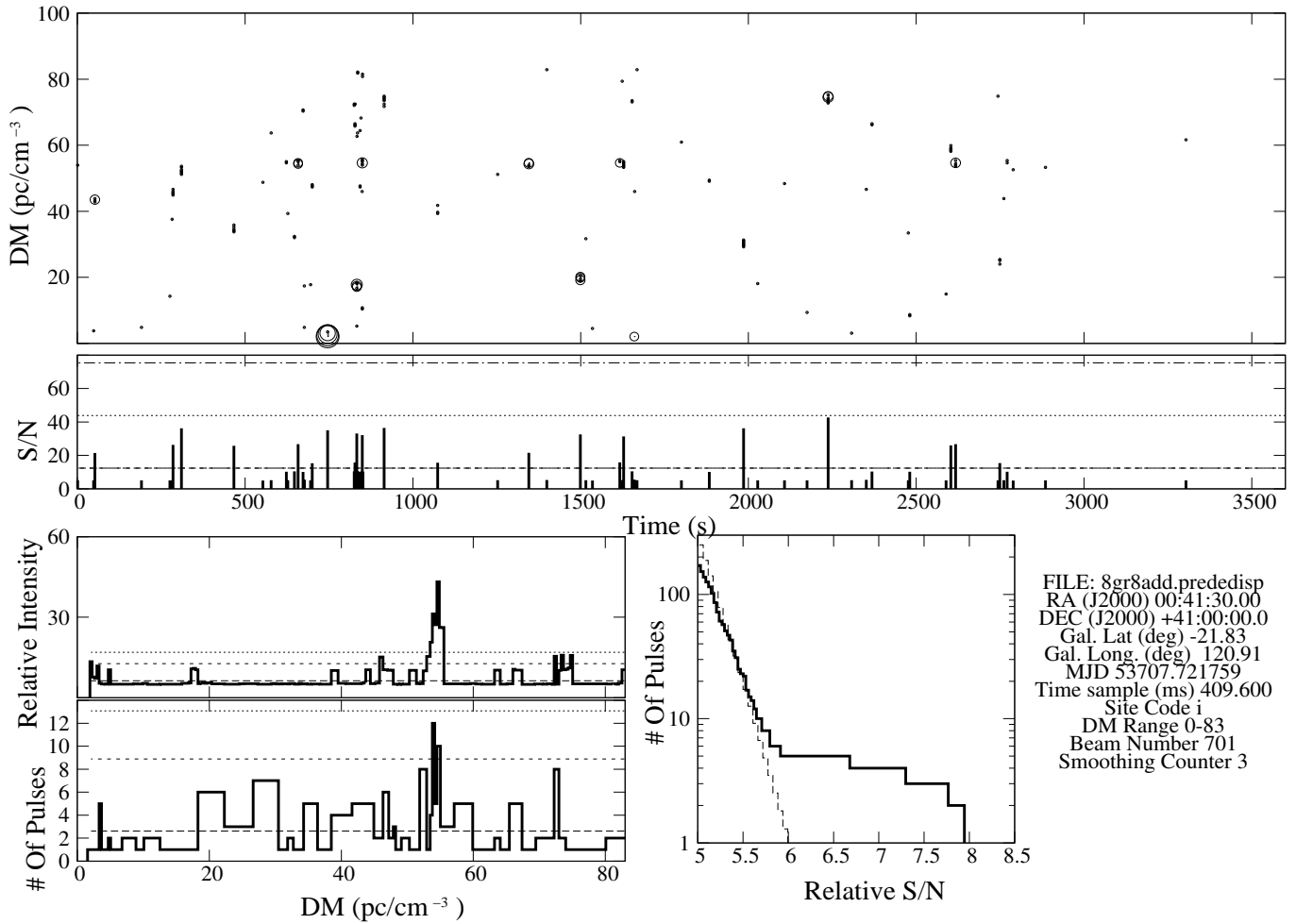}
\end{center}

\caption{Diagnostic plot from the single-pulse search, for a single
  sub-beam and a DM range $0-83$ pc cm$^{-3}$.  The top panel shows a
  DM-S/N-time array, with a time resolution of 3.2768 ms, which
    corresponds to the addition of $2^3$ raw samples of 409.6 $\mu$s.
  The data ends before t=3600 s because we analyzed a timeseries which is the
    closest power of two number of samples, which for this case is
    3436 s long. The sizes of the circles are proportional to the S/N,
  with the highest S/N $\sim$ 7.9 associated with RFI at low
  DM. Those detections that matched our RFI filter have been excised
  from the figure.  The second panel from the top shows the time series
  collapsed along DM in order to more clearly show at which times
  strong pulses occurred. The lower left panels show two
  histograms. The top one is the time series collapsed in time,
  clearly showing a peak at $\mbox{DM}\sim 54\mbox{ pc cm}^{-3}$. The
  lower one shows a histogram where time is collapsed to show the
  number of pulses versus DM. The horizontal lines in the bottom left
  panels show the mean and the 3 and 6-$\sigma$ limits of each
  distribution. The histogram in the bottom middle part of the figure
  shows the distribution of pulses with S/N $> 5$. The dashed line in
  this histogram shows the expected noise distribution.  This same
    pointing position was observed for another 31 h without recording
    the same burst profusion. \label{diagnostic-plot}}

\end{figure*}

\subsection{Periodicity search analysis }

We performed a periodicity search following the same procedure used by
\cite{jsb+05}, but adapted to the specific properties of our data. We
calculated the power spectra for each DM trial of each sub-beam using
a Fast Fourier Transform (FFT), removing frequencies known to be
related to RFI. Interpolation of the spectra was used to recover
spectral features lying between Fourier bins (\citealt{rem02}),
repeating the process after summing 2, 4, 8 and 16 harmonics.  All the
spectral features with a S/N $>$ 7 were recorded,
later keeping only the signals that were detected at multiple DMs and
appearing to be harmonically related. This was done to find
the most significant candidates, and to determine their DM and
fundamental frequency.  For each candidate, the initial period from
the spectral search was optimized by folding the data at a range of
periods, assuming a maximum error in the pulse frequency equal to the
width of one Fourier bin. No acceleration searches were done.
Typically there were a few tens of candidates per sub-beam, i.e. about
30,000 candidates per pointing.  Those combinations of DM and spin
frequency that produced the highest S/N candidates were stored for
later inspection.  This process was performed on each of the
$N_{\rm total}$ sub-beams. Once this analysis was completed, a list of
candidates from the sub-beams was collated and those candidates with
high S/N appearing in more than one sub-beam were stored. Later these
detections were compared with the lists from the other follow-up
pointings, and those candidates that matched with similar periods, DMs
and S/N were inspected by eye.

\subsection{Single pulse search analysis}

We searched for single pulses following two selection criteria:
i) single, relatively bright pulses (S/N $> 7.0$) or ii) multiple, fainter
pulses below this threshold, but repeating at the same DM. We first
discuss the general algorithm to search for single pulses, and then
discuss the details of the two selection criteria.

As mentioned in \S 3, for each sub-beam dedispersed time series are
generated at all trial DMs. Finding single pulse candidates simply
consists of recording those individual events that show an amplitude
above some threshold of S/N, which is statistically significant and
reduces the number of candidates due to RFI. An extensive description
of the method for these searches can be found in \citealt{cm+06}.

We developed a code to extract individual pulses from each of the time
series. We first determine the statistics of the noise in
the time series, which each contain $N_{\rm samp}$ samples. This is done by
calculating the mean and the rms using an iterative process where the
strongest pulses are removed until one reaches the expected number of
spurious detections ($N_{\rm noise}$) for a Gaussian distribution with
probability $P(x)$, given by $N_{\rm noise}=N_{\rm samp} P(>5\sigma)$, and
then searching again for pulses above a 5$\sigma$ threshold.  This process
keeps the mean and the rms from being biased towards very strong
individual pulses.  The S/N of a detection in the time series depends
on the intrinsic width of the pulse $W$ as S/N $\propto W^{-1/2}$,
where the pulses with the highest peak S/N will be those with a width
similar to the time resolution of the timeseries. Therefore we
explored a range of widths, such that $W_{\rm i} \geq t_{\rm s}$, by adding
between 2$^0$ and 2$^{7}$ adjacent samples, corresponding to a range
in time resolution of 0.4096 -- 52.4288 ms. Each of these smoothed
time series was then searched for single bright pulses, recording all
the pulses that showed a peak S/N $\geq 5$. For each pulse detected we
record the DM, the arrival time of the pulse relative to the start of
the observation and its S/N. Hereafter we will call this information
DM--S/N--time arrays.  The S/N threshold was chosen based on the
expected number of false detections $N (<$ threshold$)$ above that
minimum S/N by assuming that the noise is purely Gaussian.  For a time
series the number of expected false detections, above a given S/N, can
be calculated using a relation shown by \cite{cm+06}:

\begin{equation}
N (> \mbox{threshold}) \approx 2 N_{\mbox{samp}} \int_{\mbox{S/N}_{min}}^{\infty} \exp \left(-x^2/2 \right)dx.
\end{equation}
\noindent
For a full-resolution, 1-h data set, we expect $N(> 5\sigma)\sim$2700
false detections from all the time series required to cover the 549 DM
trials.\\

After searching the dedispersed time series for pulses above the S/N
threshold, we created diagnostic plots like the
one shown in Fig.~\ref{diagnostic-plot} to better assess
whether any of the signals were astronomical in origin.  Such a plot
was made for each of the sub-beams.  When sorting/inspecting the
diagnostic plots, we searched for both i) single bright pulses (S/N
$> 7.0$) and ii) multiple, weaker pulses recurring at the same DM.
As can be seen in Fig~\ref{diagnostic-plot}, the signature of a single
bright pulse appears in the top panel of the figure as a large
circle (e.g., one can see RFI between $t=500$ and $t=1000$ s), while the
signature of multiple weak pulses at the same DM appears as a
peak (in this case at $\mbox{DM}= 54 \mbox{ pc cm}^{-3}$) in either of
the two histograms in the lower left area of the plot.

\subsection{Single bright pulse search}

We recorded all the single pulses with a S/N $>5$ in the DM--S/N--time
arrays for each of the smoothed time series, keeping only the
information corresponding to those pulses that were detected in a
reasonable number of beams ($ 1/8N_{\rm total} > N_{\rm b} > 1\%N_{\rm total}$).
For each candidate, we recorded the DM, S/N, time of the event, the
number of sub-beams in which the pulse was detected, and in which
sub-beam the detection was strongest.  We then searched for possible
associations, i.e. other pulses occurring at the same DM at later
times or in other observations. Those events that had high S/N ($>
7.0$) but that did not show any associations were classified as single
bright pulses.

\subsection{Multiple pulses at the same DM}

To search for multiple, potentially weaker events occurring at the
same DM we performed a statistical analysis, searching for any excess
of S/N in the S/N-DM space.  Any excess of S/N due to real pulses
would create peaks like those shown in the lower left panels of Figure
\ref{diagnostic-plot}. Detections showing features of this kind, and
appearing in a reasonable number of sub-beams ($N_{\rm b} > 1\%N_{\rm total}$)
were recorded in a list for further inspection.  Since there are
900--1100 sub-beams per hour of observation, this produced a very
large number of candidates. In order to reduce the number of
candidates to inspect, we filtered the list by searching each
candidate for ``adjacent friends'', i.e. other detections located at
adjacent DM trials and occurring at the same time.  \cite{cm+06}
quantify how the S/N decreases when a pulse is detected at an
incorrect DM.  Narrow pulses will be detected in few DM bins,
therefore will show few adjacent friends, while broader pulses are
detected along many DM trials.  An example event with adjacent friends
is shown in the lower left panels of Fig. \ref{diagnostic-plot}. This
criteria was implemented to estimate the number of adjacent friends a
signal of a given DM should have, given the S/N of two adjacent
detections.  Once the data was filtered in this way, the remaining
diagnostic plots were inspected by eye.

\subsection{Radio frequency interference excision}

To excise RFI from the single-pulse diagnostic plots, we inspected the
DM--S/N--time arrays of each of the 8 grating groups.  Events
aligned in time and appearing along all trial DMs in all 8 data sets
were flagged as RFI and recorded in a mask file, in order to excise
those time bins when the single pulse analysis was performed.

\begin{figure}
\hspace{-0.50cm}
\includegraphics[width=0.4\textwidth,height=0.5\textwidth,angle=-90]{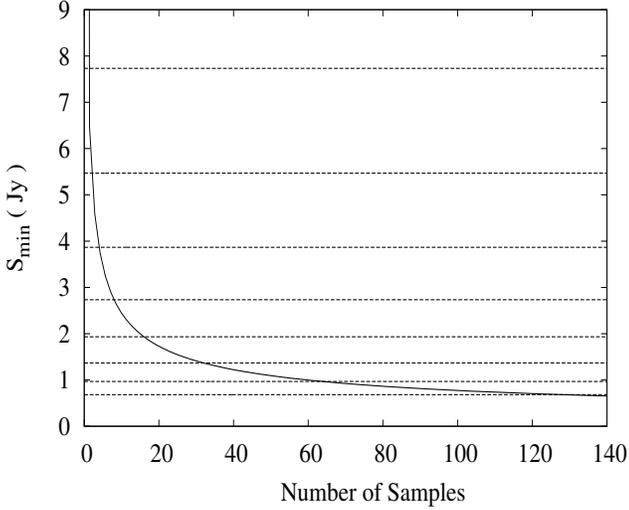}

\caption{Minimum detectable flux density for single bright pulses as a
  function of the number of samples added, assuming a constant peak
  amplitude, so that sensitivity appears to improve with
    increasing width. The raw sampling time of our data is
  $t_{\rm s}=409.6$ $\mu$s and a minimum threshold of S/N=5 was used.
  Each horizontal line represents the minimum detectable flux for each
  smoothed time series adding from top to bottom, 2$^0$ to 2$^7$
  samples.\label{minimum-sn}}

\end{figure}

\subsection{Sensitivity of WSRT in {\it 8gr8} mode }

For periodicity searches, the minimum detectable flux density is given
by the modified radiometer equation as follows (\citealt{kl05}):

\begin{equation}
 S_{\rm min}= \beta \frac{\mbox{S/N}_{\rm min} T_{\rm sys}}{G \sqrt{n_{\rm p} t_{\rm d} \Delta f}}\sqrt{\frac{W}{P-W}}.
\end{equation}
\noindent
where $\beta$ is an efficiency factor $\sim 1.3$ related to the 2-bit
sampling. S/N$_{\rm min}$ is the detection threshold, $T_{\rm sys}$ is the
system temperature in K ($T_{\rm sky}+T_{\rm receiver}$), $G$ is the gain of
the telescope (in K Jy$^{-1}$), $n_{\rm p}$ is the number of polarizations,
$\Delta f$ is the bandwidth, $P$ is the pulsar period and $W$ is the
intrinsic width of the pulse. Using $T_{\rm sys}/G= 140$ Jy, $\Delta f=10$
MHz, $n_{\rm p}=2$, a dwell time of $t_{\rm d}=$3600 s, and assuming a mean duty
cycle of 5\%, we can reach a minimum flux density of $\sim$0.25 mJy
for a S/N=7 detection.

Similarly, for the single pulse searches, our sensitivity is given by
the radiometer equation, assuming matched filtering to a pulse that is
temporally resolved, as follows (\citealt{cm+06}):

\begin{equation}
S_{\rm min} =  \frac{\mbox{S/N}_{\rm min}  T_{\rm sys}}{G \sqrt{n_{\rm p}  \Delta f  W}}.
\end{equation}
\noindent
The minimum detectable flux density depends on the pulse width as
$W^{-1/2}$, assuming a constant peak amplitude, so the sensitivity
appears to improve with increasing width.  As described in \S 3.2, we smoothed the data
before recording the peak fluxes in order to detect pulses of
different widths. The sensitivity curves for each smoothing are shown
in Fig. \ref{minimum-sn} for a threshold S/N$_{\rm min}$=5. It is
important to note that the sensitivity decreases for broader pulses
because we have effectively added more bins above the threshold.

\section{Results}

\begin{figure}
\begin{center}
\hspace*{-0.5cm}
\includegraphics[height=9cm,width=6.5cm,angle=-90]{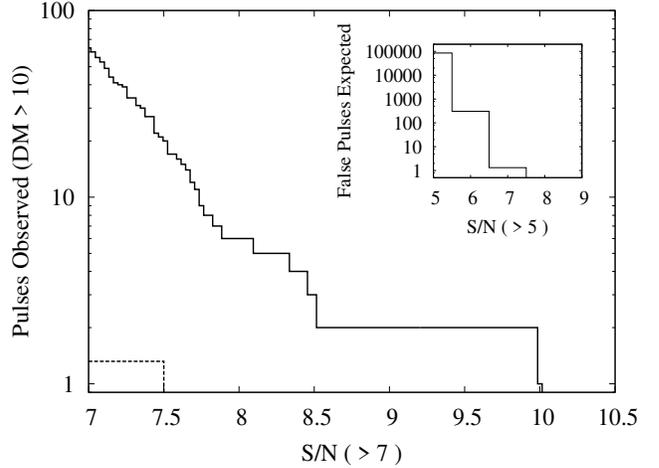}
\end{center}

{\caption{The main panel shows only detections recorded with a DM $>$ 10 pc
    cm$^{-3}$ (to avoid low-DM RFI), following the procedures described in \S3.3. All the
    pulses presented here appear dispersed, were detected in many
    sub-beams, and showed adjacent friends. The inset shows the number
    of expected false detections for purely Gaussian noise, i.e. not
    including detections due to RFI. The tail of this distribution is
    shown with the dashed line in the lower left corner of the main
    figure. Note the vertical logarithmic
    scale.  \label{histogram-SN}}}

\end{figure}

\subsection{Periodicity searches}

As described in \S 3.1, we performed a periodicity search on each 1-h
chunk of data.  No periodic sources were found using this method. We
note that there are no known foreground pulsars in the direction of
our observations. The closest known pulsar is PSR B0053+47, located
about 7.11$^{\circ}$ away from the center of our nearest pointing. In this
search, the minimum detectable period for a DM $\sim$ 50 pc cm$^{-3}$
pulsar is roughly a few milliseconds.

\subsection{Single pulse searches}

The single-pulse search produced a number of interesting
candidates. As explained in \S 3.3 and \S 3.4 we sifted our
single-pulse candidates to look for both bright individual pulses (S/N
$> 7.0$) and for multiple, weaker pulses occurring at the same DM.  We
discuss the results of each sifting method below. In
Figure~\ref{histogram-SN} we show a histogram with the distribution of
the S/N for the brightest pulses reported here. All these pulses were
detected according to the methodology described in \S 2. The excess of
detections above $7\sigma$ that match our selection criteria, and that
are shown in this plot implies that these pulses may have an
astrophysical origin in the absence of RFI or instrument related
noise.

\subsubsection{Single bright pulses}

We have detected a number of signals, not obviously associated with
RFI, with S/N $>7$, DM between $10 < \mbox{DM} < 315 \mbox{ pc
  cm}^{-3}$, and pulse widths of $1 < \mbox{W} < 27 \mbox{ ms}$, 
  as calculated from the (down-sampled) time series they were detected
  in. We present a table of these detections, as well as
corresponding diagnostic plots for selected candidates with high DM,
in Appendix A.

\begin{table}
\centering
\caption{Summary of the pulses shown in Figure \ref{diagnostic-plot}.  These pulses all peak in S/N at $\mbox{DM}=54.7$ pc cm$^{-3}$ and were detected in the same smoothed time series (pulse width of $\sim$ 3.2 ms). The arrival times are relative to MJD 53707.72175926 (topocentric). \label{tableDM55} }
\begin{tabular}{l r r r }
\hline
\hline
DM   & S/N  & Arrival Times  & \# Sub-Beams   \\
(pc cm$^{-3}$)         &      & (s)  &            \\
\hline
54.7  & 5.1 &  623.0975       &    49        \\
54.7  & 5.5 &  658.2706       &    13    \\
54.7  & 5.8 &  849.0459       &    24    \\
54.7  & 5.7 & 1345.6613       &   121    \\
54.7  & 5.5 & 1616.5904       &    22    \\
54.7  & 5.8 & 2617.5119       &    53    \\
\hline
\hline
\end{tabular}
\end{table}

\begin{figure}
\hspace*{-1mm}
\includegraphics[height=0.45\textwidth,width=0.35\textwidth,angle=-90]{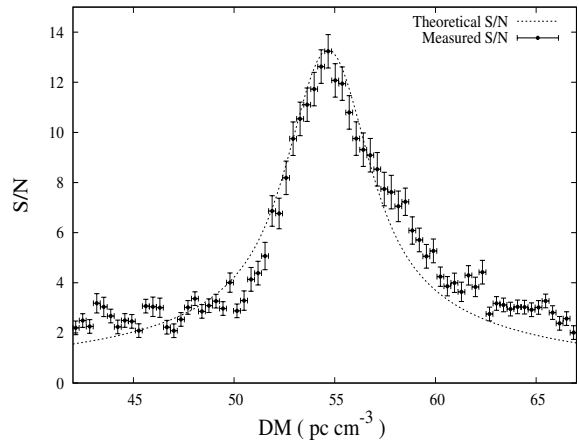}

{\caption{Summed S/N of the 6 pulses shown in Table \ref{tableDM55},
    plotted over the theoretical S/N curve for a dispersed radio pulse
    with these properties and observing parameters. The error bars in
    the $x$--axis correspond to half the DM spacing while the errors
    in the $y$--axis were calculated from the differential form of
    Eq. (3). \label{summed-sn-dm}}}

\end{figure}

\subsubsection{Multiple pulses at the same DM}

Our search for multiple pulses at the same DM revealed a few events
that satisfy the conditions described in \S 3.3. The most convincing
detection is shown in Fig.~\ref{diagnostic-plot}, in which 6 weak
bursts ($5.1 \leq$ S/N $\leq5.8$) were recorded during a single hour.
The details of each of these detections are listed in
Table~\ref{tableDM55}. This is the most tantalizing detection made in
our survey; we did not observe any other candidate with the same burst
profusion.  At this DM, other pulses were detected in the remaining 31
h of data, but these occurred only sporadically. 

For a Gaussian noise distribution, and using the dwell
times of our observations, the sampling time and the number of beams,
we estimate that the probability of having a $\sim 5.5\sigma$
detection, with adjacent friends is about $< 10^{-7}$ and that the
probability of having 6 bursts with adjacent friends is much lower
($\ll 10^{-7}$).

Using the PRESTO tool {\tt rrat\_period}, which performs a brute-force
period search given a list of pulse arrival times, we determined two
candidate periods based on the burst times listed in
Table~\ref{tableDM55}.  This analysis assumes that the pulse period is
$> 50$ ms.  If the actual period is less than this, then the time gap
between the pulses is too large to unambiguously determine a period.
Using all 6 arrival times, the best candidate period is 0.23294 s,
implying the following number of rotations between the 6 arrival
times: 150.9977, 818.9964, 2131.9647, 1163.09577, and 4296.9455.
Dropping the first, weakest burst, the best candidate period is
0.29578 s, implying the following number of rotations between the 5
arrival times: 644.9923, 1679.0073, 915.9843, and 3384.0160.  To
    check how robust these candidate periods are, we performed
    numerous trials in which we adjusted the pulse arrival times to
    arrive randomly within a 10-second window after the true arrival
    time.  We then re-ran {\tt rrat\_period}.  Most of these trials
    resulted in candidate periods that fit the randomized arrival
    times just as well as for the real data.  Thus, we caution that
    the candidate periods we present here do not conclusively indicate
    that these bursts have an underlying periodicity.

\begin{figure}
\begin{center}
\hspace*{0.5cm}
\includegraphics[height=0.375\textwidth,width=0.35\textwidth,angle=-90]{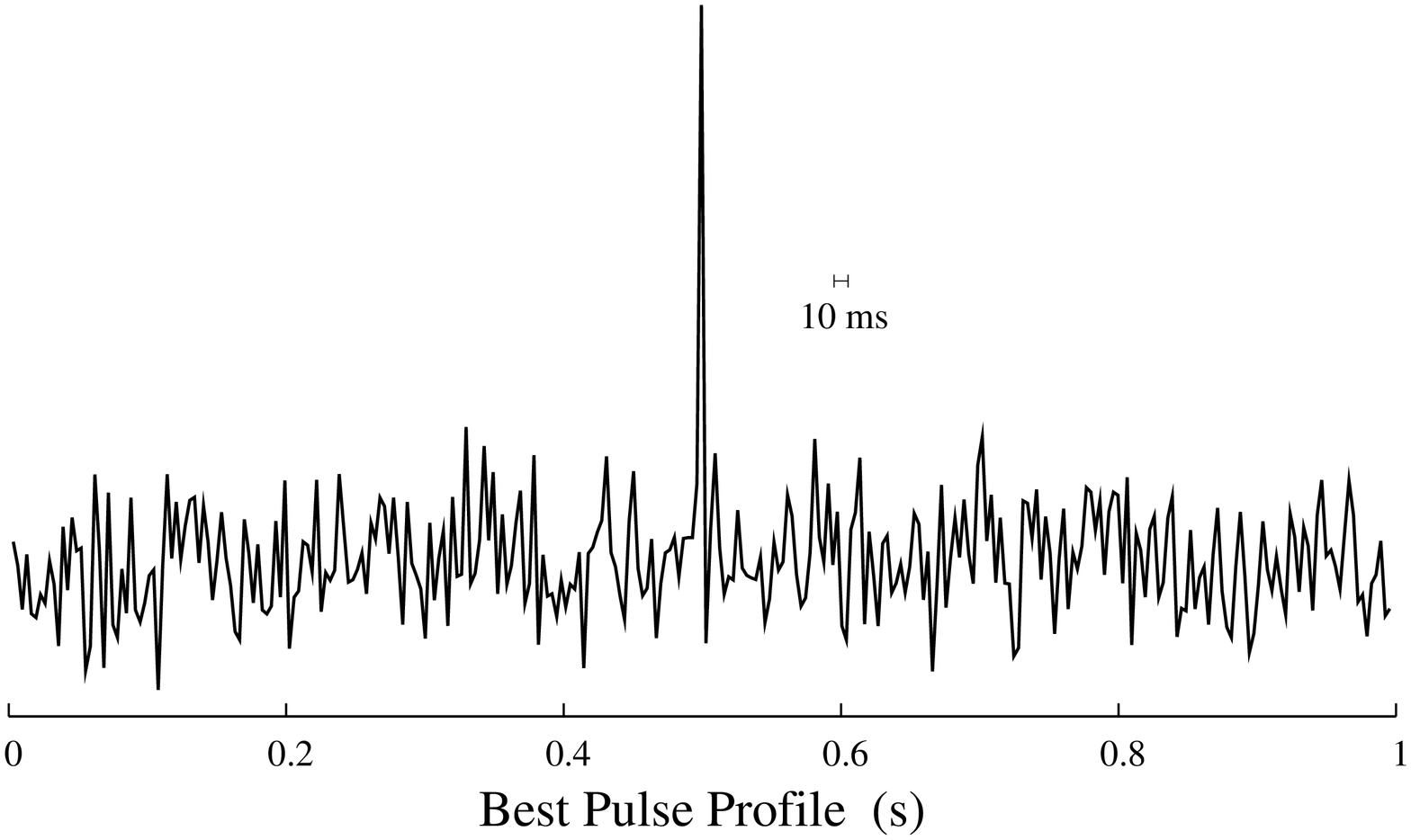}

\vspace*{-2.cm}
\hspace*{-2cm}
\includegraphics[height=0.75\textwidth,width=0.55\textwidth,angle=-90]{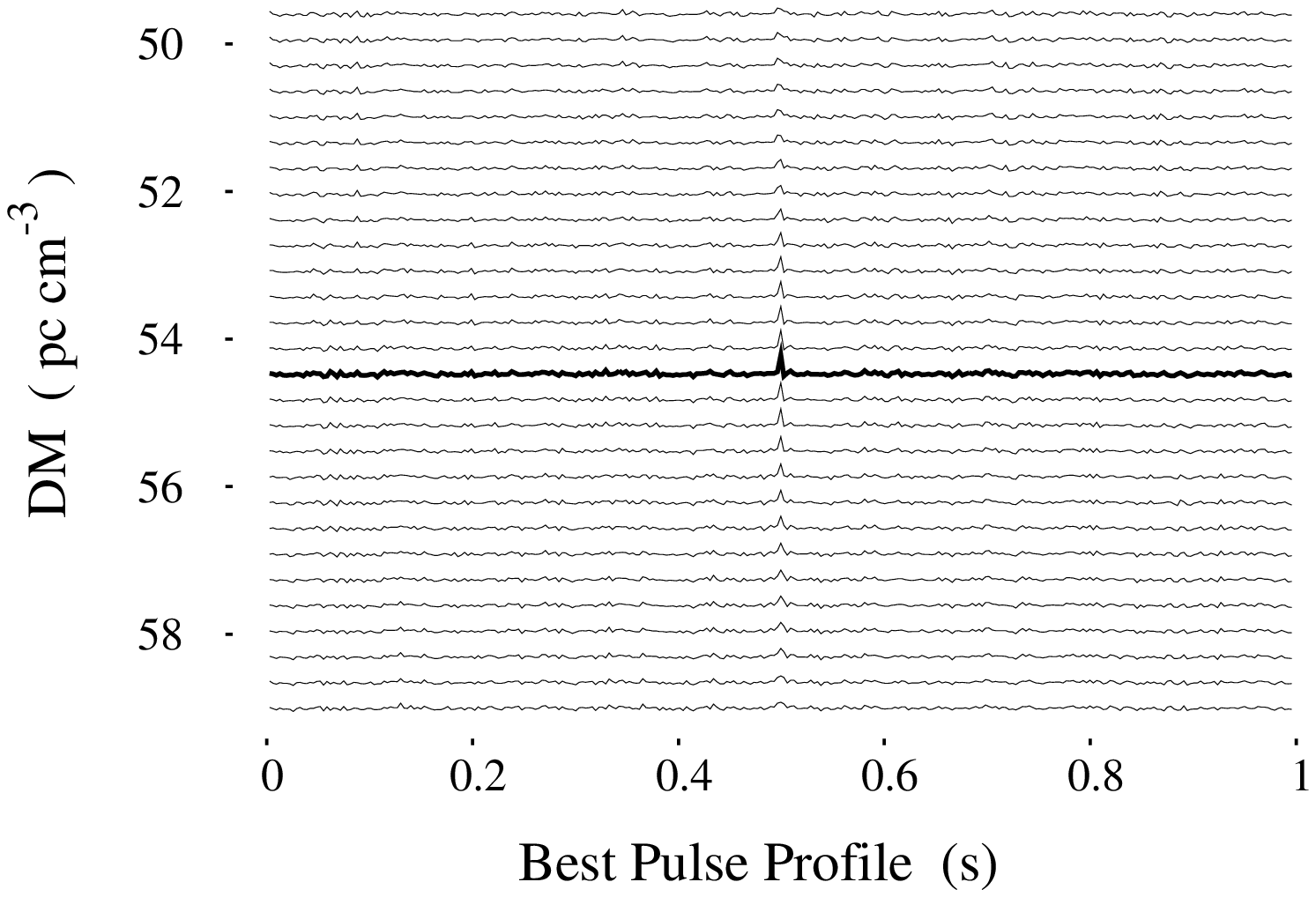}
\vspace*{-1.5cm}

{\caption{Summed pulse profile of our best candidate. The upper panel
    shows the pulse profile obtained after adding the 6 pulses listed
    in Table \ref{tableDM55}. The lower panel shows how this summed
    pulse widens and decreases in peak S/N at adjacent DM trials
    both above and below DM=54.67 pc cm$^{-3}$ (thick line).  Due to the very low S/N of the individual pulses, a
      frequency-time waterfall plot showing the dispersion
      sweep is not informative.\label{waterfall-plot}}}

\end{center}
\end{figure}

To better test whether these bursts are indeed astronomical in nature,
we compared the measured S/N of the sum of all 6 pulses, as a function
of DM, with the expected theoretical curve (following Eqs. 12 \& 13
given in \citealt{cm+06}).  Figure \ref{summed-sn-dm} shows that the
summed signals match the theoretical curve well.  To make this summed
profile, we added the time series of each pulse, using the peak of each
burst as the fiducial alignment point.  The resulting best pulse
profile is shown in the top panel of Fig.~\ref{waterfall-plot}. From
this, we can conclude that the pulse width is comparable or shorter
than the sampling time of these time series, i.e. $\leq$ 3.2 ms.  The
bottom panel of Fig.~\ref{waterfall-plot} shows the time series of
adjacent DM trials, after being corrected for dispersion and spaced in
DM as in our original search (see \S3). The figure clearly shows the
dispersed nature of these bursts.

In total we have detected 31 other candidates where weak pulses
repeated with a lower burst rate than the candidate described
above.  From these 31 we report here those with the highest burst
rate. Their characteristics are shown in Table
\ref{other-multiple-DM-detections}. These sources have been detected
throughout the area covered by our observations and some were found
where our observations overlap (see Fig. \ref{m31-beams}). They were
identified as promising because of a significant overabundance of
pulses at a specific DM. The S/N of these pulses range from $5 <$ S/N
$<7$. The typical burst rate detected with the WSRT for these
candidates was between 1 -- 3 per hour.  The DMs of these detections
are generally consistent with that of an object located at least in
the outskirts of our Galaxy and potentially in M31. For each of the
sources listed in Table~\ref{other-multiple-DM-detections} we
calculated the most likely location within the primary beam by combining the data from the individual pulses with the
highest S/N in each observation.  In Table~\ref{results_single_pulse2}
of Appendix B we present the brightest multiple pulse detections we
recorded. These events repeated more than once at the same DM and with
a similar pulse width.

\begin{table*}
\begin{minipage}{156mm}

\caption{The positions of our best four candidates detected using
    the criteria described in \S4.2.2 for multiple pulses detected at
    the same DM.  For each candidate we list the right ascension (RA)
  and declination (DEC) of the best positions (found using the method
  described in \S2), the number of bursts detected per hour $\dot{N}$
  (incorporating the full data set), the DM, the inferred
  distance (d) using the NE2001 Galactic free electron model and the
  width of the individual pulses $W$ corresponding to the time
    resolution at which the pulses had maximum significance in the search.  The
  errors in position for RA can be up to 0.5 arcmin while in DEC they
  are 1.5 arcmin. The numbers in parentheses represent the uncertainty
  on the least significant digit in the
  DM. \label{other-multiple-DM-detections}}

\centering
\begin{tabular}{l c c c c c c}
\hline
\hline
 Name       & RA(J2000) & DEC (J2000)          &  $\dot{\mbox{N}}$   &  DM~~~          &  d               &    $W$         \\
            & h~~~m     & d~~~m                   &  h$^{-1}$     &  pc cm$^{-3}$   & ~~kpc            & ms        \\
\hline
  M31--BC1  & 00 43     & +40 22     & 2.7            & $54 (1)$       & $\geq$ 2.9     & $\sim$ 3.3  \\
  --        & 00 44     & +40 39     & --              &  --            &  --               &  --         \\
  --        & 00 43     & +40 58     & --              &  --            &  --               &  --         \\
            & 00 46     & +41 26     & --              &  --            &  --               &  --         \\         
            & 00 44     & +41 44     & --              &  --            &  --               &  --         \\  
\hline
  M31--BC2  &  00 39    & +41 38     & 2.7            & $56 (1)$       & $\geq$ 3.2     & $\sim$ 3.3  \\
  --        &  00 39    & +41 14     & --              &  --            &  --               &  --         \\
  --        &  00 39    & +40 30     & --              &  --            &  --               &  --         \\
\hline
  M31--BC3  &  00 37    & +40 24     & 3.0            & $63 (1)$       & $>$ 48       & $\sim$ 3.3   \\ 
  --        &  00 47    & +40 53     & --              &  --            &  --               &  --         \\
  --        &  00 40    & +40 40     & --              &  --            &  --               &  --         \\
\hline
  M31--BC4  & 00 45     & +40 54     & 1.0           & $147 (2)$      & $>$ 48        & $\sim$ 6.6   \\
  --        & 00 43     & +40 45     & --              &  --            &  --               &  --         \\
  \hline
  \hline
\end{tabular}
\end{minipage}
\end{table*}

\subsubsection{Localization of the candidates}

\begin{figure}
\hspace*{-1mm}
\includegraphics[height=0.45\textwidth,width=0.5\textwidth,angle=0]{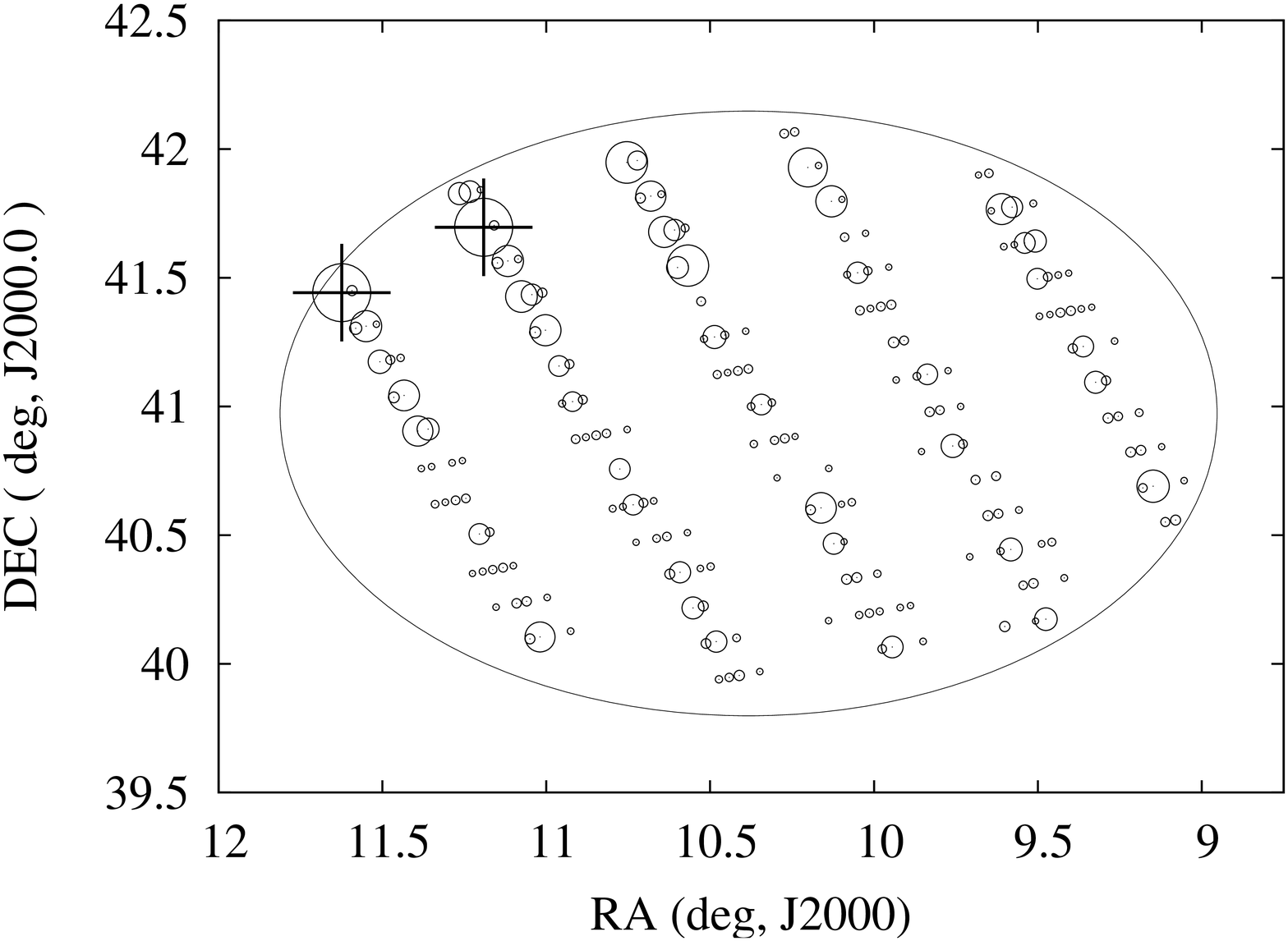}

{\caption{Beam plot of the bursts detected at $\mbox{DM}= 54.7$ pc
    cm$^{-3}$. The size of the circles is proportional to the S/N of
    the combined detections. The highest S/N regions are located in
    the upper left side of the plot (see crosses) and are
    $\mbox{RA}=11.62 \degr$, $\mbox{DEC}=+41.44\degr$ and
    $\mbox{RA}=11.19\degr$, $\mbox{DEC}=+41.69\degr$ respectively. The
    large circle represents the approximate size of the primary
    beam. \label{beam-plot-DM55}}}

\end{figure}

Due to the sporadic nature of the single pulses, and the linear nature
of the WSRT array, we were not able to unambiguously locate any of the
sources reported here within the primary beams.  To unambiguously
locate a source, we require at least a handful of bright single pulse
detections at different hour angles, as explained in \S 2.  We show an
example of this procedure for the pulses of the candidate M31-BC1 (see
Table \ref{other-multiple-DM-detections}). Combining only the S/N of the
detections listed in Table \ref{tableDM55} we obtain two best
locations for this source which are shown in Figure
\ref{beam-plot-DM55}. Applying this procedure and includding all the detections, 
we find the best locations for the candidates described in \S4.2.2 and listed in
Table~\ref{other-multiple-DM-detections}.

\begin{figure}
\hspace*{-1mm}
\includegraphics[height=0.45\textwidth,width=0.5\textwidth,angle=0]{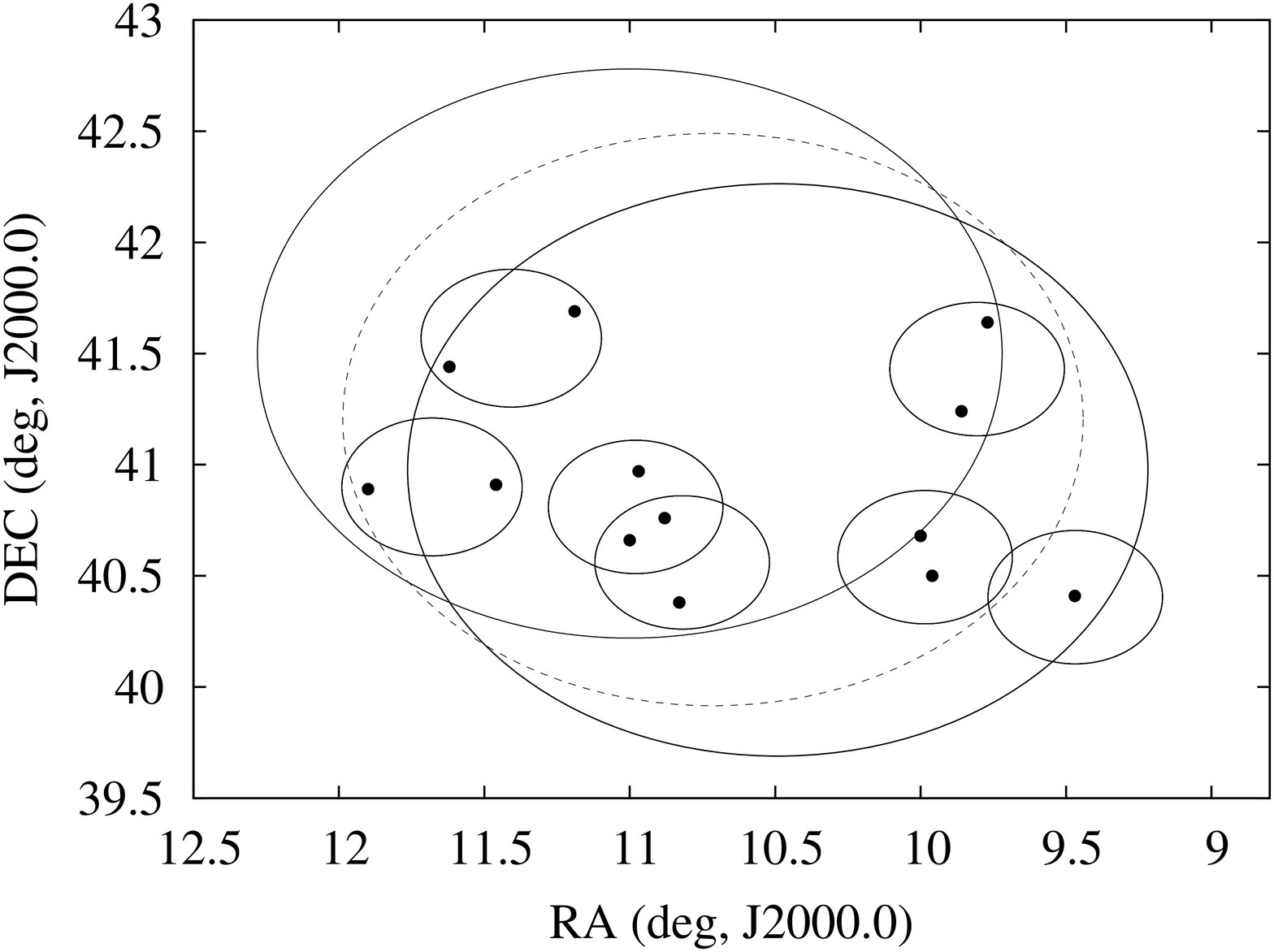}

{\caption{The medium-sized circles show the area covered by
    follow-up GBT observations.  The FoVs of the original WSRT observations are
    represented by the large circles. The black dots show the position
    of the candidates listed in Table
    \ref{other-multiple-DM-detections}. \label{pointings-gbt}}}

\end{figure}

\subsubsection{Follow-up observations}

The candidates shown in Table~\ref{other-multiple-DM-detections} were
followed up first by observations using the standard WSRT tied-array
mode. These were performed at the same observing frequency and
using the same number of frequency channels as in our {\tt 8gr8}
observations. At this frequency, the WSRT tied-array beam has a radius
of $\sim3^{\prime}$, about half the size of an {\tt 8gr8}
sub-beam. Each position was observed for approximately 2 h. These
observations did not confirm any of the candidates. The low S/N of
most of these detections makes differentiation between real pulses and
spurious (noise-related) pulses very difficult, even though the pulses
appear dispersed and are recorded in multiple sub-beams. This might
introduce systematic errors in the inferred positions that are larger
than the {\tt 8gr8} sub-beams, and hence it is entirely possible
  that we were simply not pointing in the right direction during the confirmation attempts.

In order to circumvent this problem, we decided to attempt
confirmations with an instrument that provides larger beam size and
greater sensitivity: the Green Bank Telescope (GBT) and Pulsar Spigot backend
(\citealt{kel+07}) at a center
frequency of 350 MHz. Each of the best positions was observed for 2 h,
for a total campaign time of 16 h.  The pointings observed with the
GBT are shown in Figure \ref{pointings-gbt}. We used a bandwidth
$\Delta f=50$ MHz with 1024 lags and a sampling time of 81.92
$\mu$s. The GBT and Pulsar Spigot provided roughly a factor of two higher gain and five
times the recorded bandwidth, meaning an improvement of a factor of 4
in sensitivity over the original {\tt 8gr8} observations. At this
frequency the beam size of the GBT is $0.6^{\circ}$. We reduced these
observations using the PRESTO package~(\citealt{rem02}),
applying both periodicity and single pulse searches.  From the burst
rates cited in Table \ref{other-multiple-DM-detections}, we were
expecting to detect at least one of these candidates, but no
confirmations were made. Assuming that these candidates are
  indeed from true astronomical sources, the lack of detections can
be attributed to two reasons: i) incorrect localization due to the low
S/N of the original detections and ii) the possibility these sources
manifest in rare, sporadic clumps of bursts, requiring a longer dwell
time to detect.

\section{Discussion}

Our search of M31 can be compared with, e.g., the Arecibo 350-MHz
search of M33 by \citet{mc+06}: both searches detected some sporadic
single pulses with DMs consistent with an origin in the target galaxy.
To compare the two searches, we consider that: i) M33 is $\sim$120 kpc
more distant than M31 (\citealt{kkv+3}), making any source $1.3
\times$ fainter than it would be in M31, (ii) the Arecibo radio
telescope is 10 times more sensitive than the WSRT at these
frequencies, and (iii) the low mass, different metallicity and star
formation rate of M33 would likely produce a smaller pulsar population
than that of our own Galaxy and M31. Taking into account these
differences, it turns out that our results are similar to those found
by \cite{mc+06} for M33, as we demonstrate below.

\subsection{Detectability of  pulsars and RRATs in M31}

Here we consider the detectability of pulsars in M31 using the
periodicity and single pulse search techniques.  This, in turn, is
used to place constraints on the total pulsar population of M31. To do
this, we constructed a simple population synthesis code to model the
basic properties of a galactic pulsar and RRAT population. Using the
sensitivity thresholds listed in \S 3.6 and our synthesis code, we
model different populations to find an average expected number of
detectable pulsars and RRATs. For periodicity searches we used
integration times of 1 h, while for single pulse searches we used
observation times of 8 h.

\subsubsection{Modeling pulsars}

\begin{figure*}
\hspace{-0.40cm}
\includegraphics[width=0.4\textwidth,height=0.5\textwidth,angle=-90]{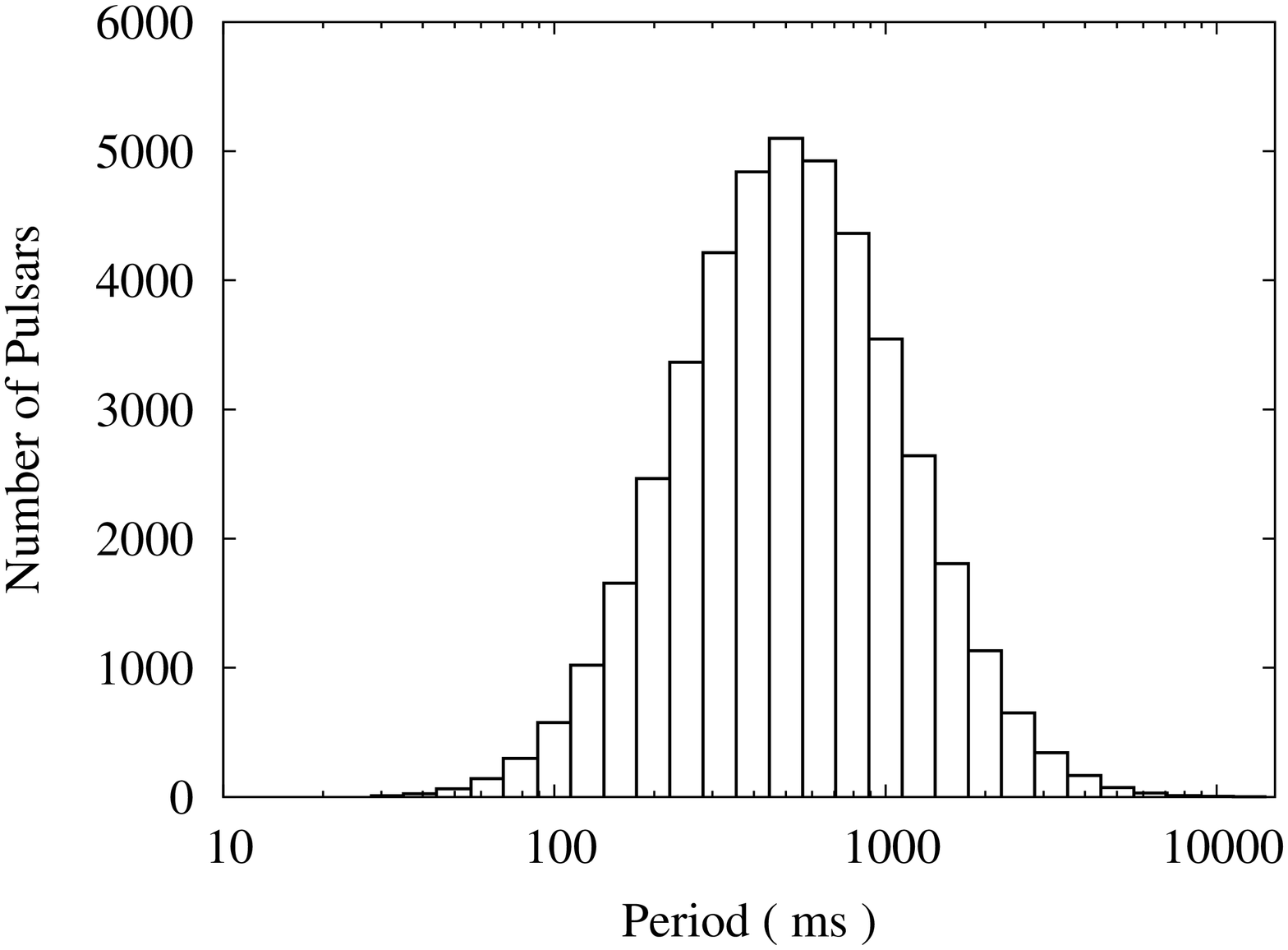}
\includegraphics[width=0.4\textwidth,height=0.5\textwidth,angle=-90]{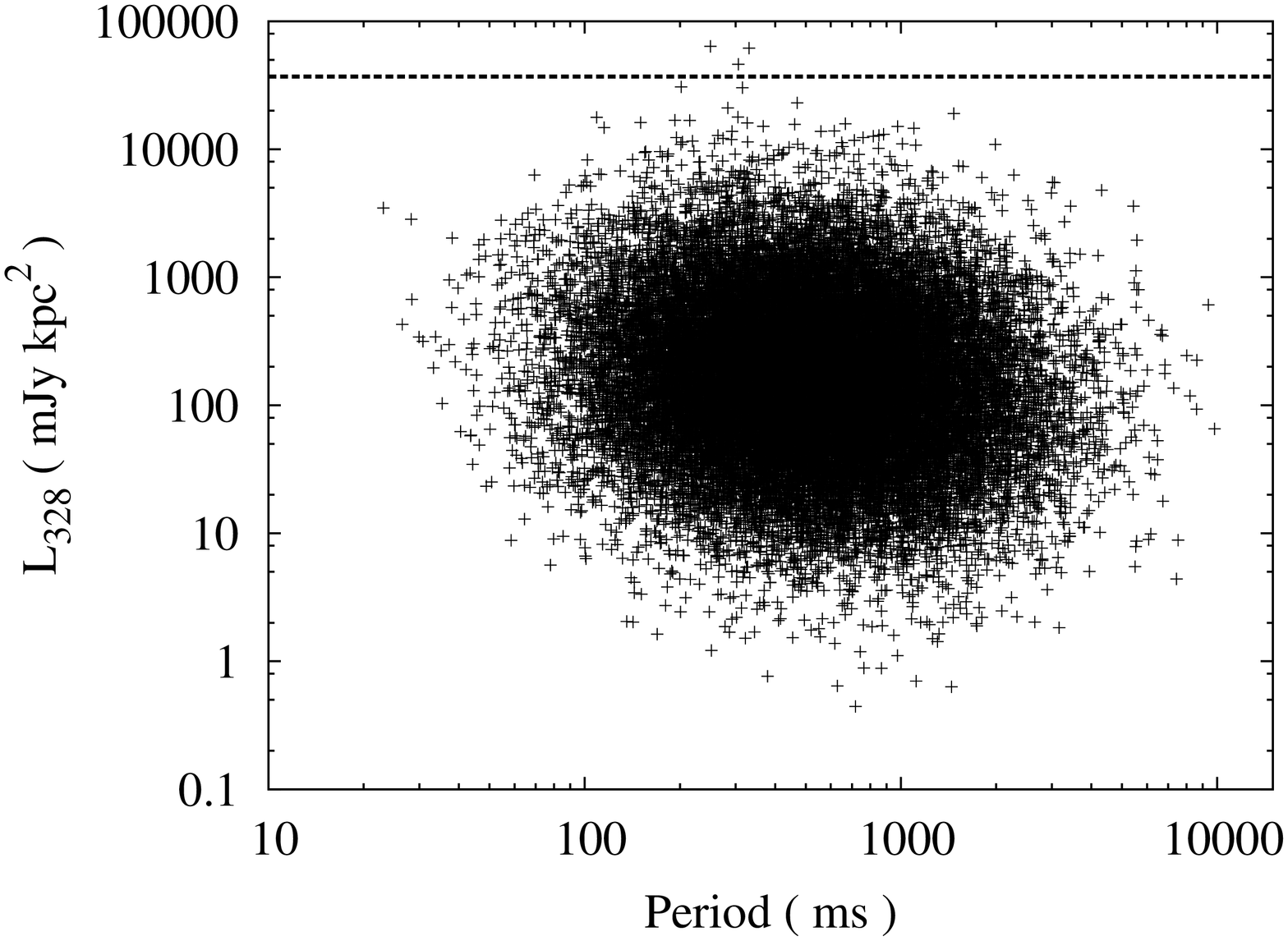}

\caption{One example model realization for radio pulsars. The left
  panel shows a histogram of the number of pulsars versus period. Our
  models cover periods ranging from a few tens of milliseconds to a
  few seconds. The right panel shows the distribution of luminosities
  versus period. The dotted line represents the most luminous known
  pulsar in our Galaxy, PSR B1302$-$64, which has a $L_{328} \sim 37$
  Jy kpc$^2$. \label{pulsar-models}}

\end{figure*}

In our pulsar population model, we simulate the distribution of pulsar
periods $N(P)$, luminosities as a function of pulse period $L(P)$, and
the distribution of pulse energies for each pulsar. Each synthesized
population contains about 40,000 pulsars, which is based on simulations
of our own Galaxy (see, for example, \citealt{lori09}). In other
words, we assume that the number of pulsars in M31 and our Galaxy is
similar.  We do not simulate millisecond pulsars because they do not
represent the bulk of the population of pulsars in our Galaxy and, in
any case, we had poor sensitivity to them. For the period distribution
we use the function found by ~\cite{lfl+11}:

\begin{equation}
N(\log P)= A \exp \left[-\frac{(\log P -B)^2}{2 C^2}  \right].
\end{equation}
\noindent
For the distribution of luminosities we derived an empirical relation
from a sample of 663 pulsars with known $S_{400}$, which we obtained
from the
ATNF\footnote{http://www.atnf.csiro.au/research/pulsar/psrcat/}
catalog (\citealt{mhth05}). We then fit a log-log distribution with a
normal distribution given by:

\begin{equation}
\log(L) = D \log(P) + E +\zeta.
\end{equation}
\noindent
Where $\zeta$ is a normal distribution centered at the origin with
standard deviation 0.6. The parameters of our population models are
summarized in Table \ref{parameters}. An example of the period and
luminosity distributions for one of our models is shown in Fig.
\ref{pulsar-models}.

For each pulsar, we generated a random pulse-energy distribution,
which was either i) normal, ii) log-normal, or iii) a power law.
These are the distributions observed in known pulsars
(e.g. \citealt{cjd01} and \citealt{wws+06}).  For power law
distributions, the index $\alpha$ was assigned randomly, with values
between $1 < \alpha < 4$. For the normal and log-normal distributions,
the mean was assumed to be the mean flux density of each pulsar
calculated from the luminosities given by Eq. (5), using the
definition of luminosity $L_{328}=S_{328}d^2$, with $d=778$ kpc
(\citealt{kkv+3}, the distance to M31). For the normal distributions,
we randomly assigned a $\sigma$ to simulate the distributions of
pulses in the range $100 > \sigma > 1$. For the log-normal
distribution, we followed the same approach, assigning randomly a
$\sigma$ that would produce a range of distributions. For each pulsar
we also calculated the number of pulses that we should observe in one
of our 8-h observations (given by $N_{\rm p}=t_{\rm i}/P$).  To build these
models, the luminosities of pulsars at 328 MHz ($L_{328}$) have been
scaled from those at 400 MHz, following the relation $S \propto
\nu^{-\alpha}$, where $S$ is the flux at 328 MHz, $\nu$ is the
observing frequency and $\alpha$ is the spectral index, assumed to be
on average $\alpha=-1.82$ from \cite{mkk+00}.

\begin{figure*}
\hspace{-0.40cm}
\includegraphics[width=0.4\textwidth,height=0.5\textwidth,angle=-90]{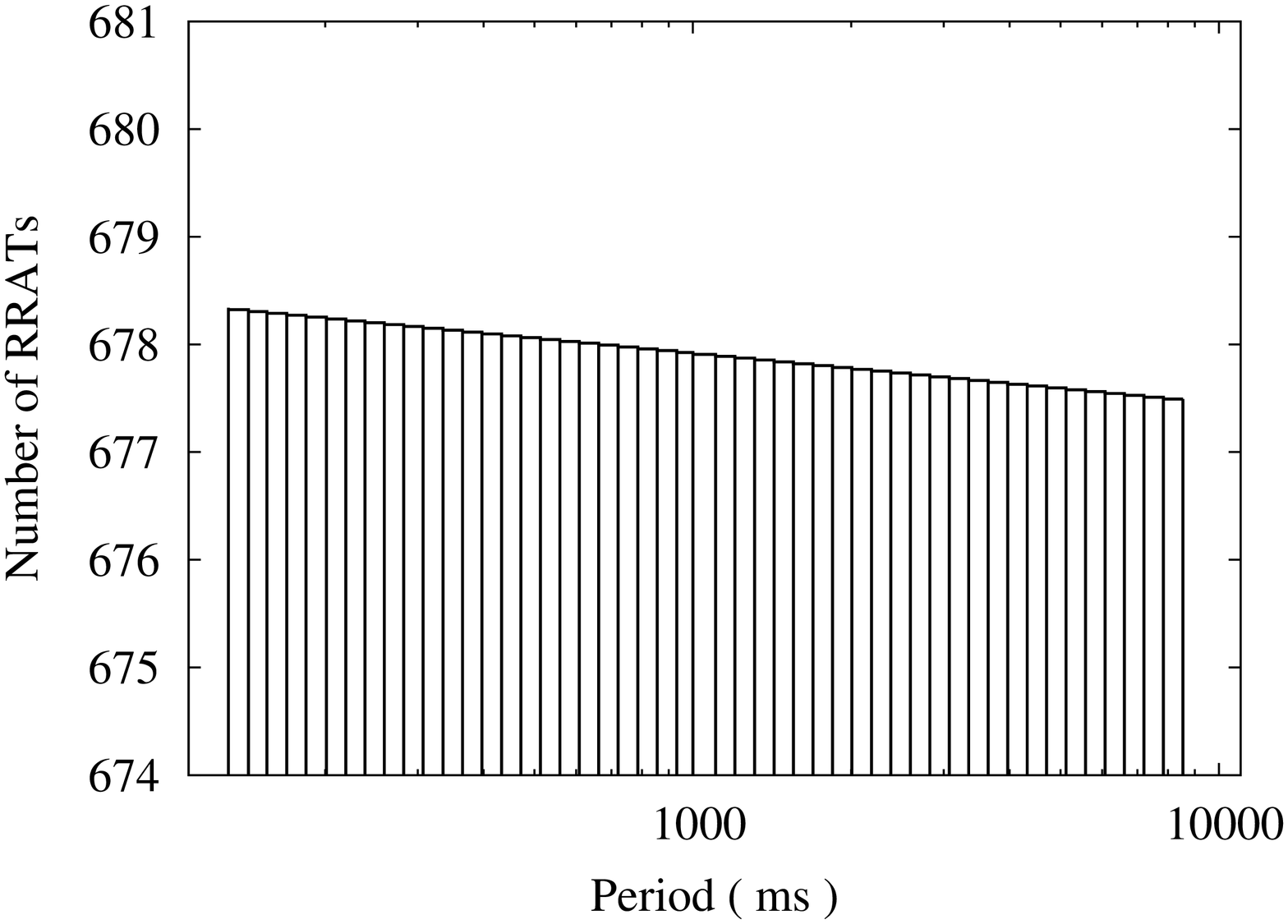}
\includegraphics[width=0.4\textwidth,height=0.5\textwidth,angle=-90]{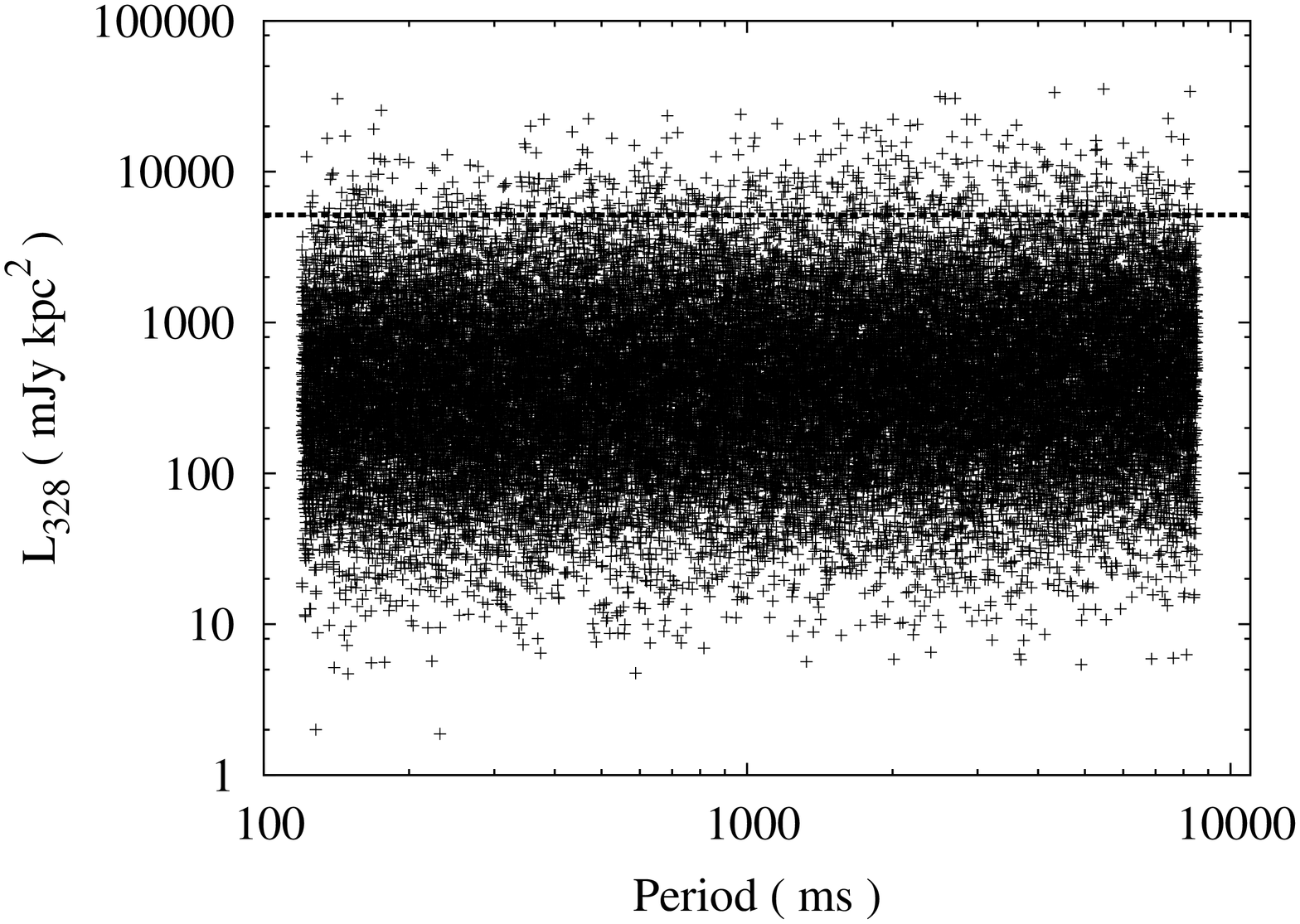}

\caption{One example model realization for RRATs.  As in
  Fig. \ref{pulsar-models}, the left panel shows the distribution of
  the number of RRATs versus period, based on the known periods of
  RRATs.  The right panel shows the distribution of luminosities
  versus period. The dotted line shows the most luminous RRAT known in
  our Galaxy, J1819$-$1458, with peak luminosity $L_{328} \sim 5170$ mJy
  kpc$^2$. The luminosities shown here were scaled with frequency as
  explained in the text. \label{RRAT-models}}

\end{figure*}

\begin{table*}
\centering

\caption{Summary of the fitted parameters from our pulsar and RRAT
  models. The various parameters used are described in Eqs. 4--8. The
  columns marked with $d^{-2}$ are the model fits using the standard
  inverse square law for flux as a function of distance. The columns
  marked with $d^{-1}$ and $d^{-1.5}$ show the fits for other assumed
  flux scaling laws, as explained in \S5.2. The spaces marked with
  ``--'' indicate that the models did not require those
  parameters.\label{parameters} }

\begin{tabular}{c r r r r r r  }
\hline
\hline
Parameter  & \multicolumn{3}{c}{Pulsar}            & \multicolumn{3}{c}{RRATs}               \\
           &  $d^{-2}$     &$d^{-1.5}$        &  $d^{-1}$    & $d^{-2}$     &$d^{-1.5}$        &  $d^{-1}$       \\ 
 \hline
A          &  0.510     &  0.510         &  0.510  & -0.467141 &  -0.467141     & -0.467141  \\
B          &  2.710     &  2.710         &  2.710  &  679.3115 &   679.3115     &  679.3115  \\
C          &  0.340     &  0.340         &  0.340  &  ---      &   ---          & ---        \\
D          & -0.230     &  -0.211798     & -0.1931 & -0.045    &   -0.0742648   & -0.0645126 \\  
E          &  3.212     &  2.36363       & -0.0645 &  2.590    &    3.47843     &  3.18352   \\  
F          &   ---      &  ---           & ---     & -0.2633   &   -0.2633      & -0.2633    \\
G          &   ---      &  ---           & ---     &  1.91755  &    1.91755     &  1.91755   \\
\hline
\hline
\end{tabular}
\end{table*}

\begin{table*}
\centering

\caption{Monte-Carlo simulations for different luminosity
  distribution scalings (shown in the first column) for pulsars. In
  the other columns we show the number of pulsars detected using the
  different methods, namely periodicity and single pulse searches. We
  did simulations using shallow ($\alpha \sim 1$) and steep ($\alpha
  \geq 2$) power law indices. For each case we show the number of
  detectable pulsars using a flux law that scales as $d^{-2}$,
  $d^{-1.5}$ and $d^{-1}$ (see text in \S5.2). The spaces marked with
  ``--'' indicate that we did not model those conditions since they
  were intermediate between the lowest and highest luminosity
  models. \label{models-pulsars}}

\begin{tabular}{ r c c c c c c c c c c c c c  }
\hline
\hline
Model        & \multicolumn{13}{c}{Average \# of Pulsars Detected}                                                                                                                \\        
\hline
\hline
             & \multicolumn{3}{c}{Periodicity Search}  & \multicolumn{10}{c}{Single Pulse }                                                                                        \\ 
             & \multicolumn{3}{c}{FFT}                 & Gaussian & \multicolumn{3}{c}{Log-Normal}  & \multicolumn{6}{c}{Power Law}                                                \\
\hline
             & \multicolumn{3}{c}{}                    &          & \multicolumn{3}{c}{}            &\multicolumn{3}{c}{$1< \alpha < 2$}   & \multicolumn{3}{c}{ $2< \alpha < 4$}  \\ 
             & $d^{-2}$&$d^{-1.5}$&  $d^{-1}$            &          & $d^{-2}$&$d^{-1.5}$&  $d^{-1}$    & $d^{-2}$&$d^{-1.5}$& $d^{-1}$          &$d^{-2}$&$d^{-1.5}$& $d^{-1}$              \\
\hline  
$L$          &   0  &     8  &   808                   &  0       & 2    & 1    & 2                 &  2   & 3       & 2                   &   0 & 5       & 0             \\
2$\times L$  &   0  &    13  &  2445                   &  0       & 2    & 2    & 2                 &  3   & 4       & 3                   &  -- & --      & --            \\
5$\times L$  &   2  &    80  &  7705                   &  0       & 2    & 2    & 3                 &  5   & 4       & 4                   &  -- & --      & --            \\
10$\times L$ &   5  &   375  & 14565                   &  0       & 2    & 2    & 2                 &  6   & 4       & 6                   &   0 & 5       & 0             \\
\hline
\hline
\end{tabular}
\end{table*}

\begin{table*}
\centering

\caption{Monte-Carlo simulations made for RRATs using different
  luminosity laws (first column). $L$ is the luminosity as seen for
  the population of RRATs as observed in our Galaxy ($L = Sd^2$). In
  the second and third columns we show the number of RRATs detected
  via their single pulses. We also show the number of detections for
  each of the flux distribution models. In the power law models, we
  did simulations using shallow ($\alpha \sim 1$) and steep ($\alpha
  \geq 2$) power law indices. As in Table \ref{models-pulsars}, the
  numbers in each column below the distributions show the expected
  number of detectable RRATs using a flux law that scales as $d^{-2}$,
  $d^{-1.5}$ and $d^{-1}$ (see text in \S.5.2).  \label{models-rrats}}

\begin{tabular}{ r c c c c c c c c c c }
\hline
\hline
Model        & \multicolumn{10}{c}{Average \# of RRATs Detected}                                                                       \\        
\hline
\hline
             & \multicolumn{10}{c}{Single Pulse }                                                                                        \\ 
             &  Gaussian & \multicolumn{3}{c}{Log-Normal}  & \multicolumn{6}{c}{Power Law}                                               \\
\hline
             &           & \multicolumn{3}{c}{}            &\multicolumn{3}{c}{$1< \alpha < 2$}   & \multicolumn{3}{c}{ $2< \alpha < 4$}  \\ 
             &           & $d^{-2}$&$d^{-1.5}$&  $d^{-1}$         & $d^{-2}$&$d^{-1.5}$& $d^{-1}$          &$d^{-2}$&$d^{-1.5}$& $d^{-1}$                    \\
\hline   
$L$          &   0       &   0  &     0  &  0              &  0   & 0      &  0                    & 0   &  0     & 0                      \\
2$\times L$  &   0       &   0  &     0  &  1              &  1   & 0      &  0                    & --  &  --    & --                     \\
5$\times L$  &   0       &   1  &     0  &  3              &  0   & 0      &  1                    & --  &  --    & --                     \\
10$\times L$ &   0       &   1  &     0  &  3              &  0   & 0      &  1                    & 0   &  0     & 2                     \\
\hline
\hline
\end{tabular}
\end{table*}

\subsubsection{Modeling RRATs}

RRATs have been relatively recently identified as a class of
  radio-emitting neutron stars and currently a few tens of these
objects are known (see the
RRATalog\footnote{http://www.as.wvu.edu/$\sim$pulsar/rratalog/}). In the
discovery paper, \cite{mll+06} attempt to simulate the size of the
population based on the 11 sources they report. Nevertheless, no
synthesis models have been made to describe in detail their individual
characteristics since our current understanding of the population as a
whole is uncertain. To build our model, we use the measured periods
and other parameters of more than 50 known RRATs reported by
\cite{mll+06}, ~\cite{hrk+08}, ~\cite{dcm+23}, ~\cite{kle+4} and
\cite{b-sb}. With this empirical model, we aim to describe the general
properties of a galactic population of about 40,000 RRATs, in order to
estimate the distribution of periods $N(P)$, luminosities $L(P)$ and
pulse rate $\dot{N}$. For the period distribution we use the following
analytic function:

\begin{equation}
N(P) = A \log(P)+B,
\end{equation}
\noindent
while for the luminosity we used the following relation:

\begin{equation}
\log L=D \log(P)+E+\Gamma.
\end{equation}
\noindent
with $\Gamma$ being a normal distribution with zero mean and standard
deviation of 0.4 (obtained from our best fits). 
The sporadic nature of RRATs can be quantified by
the pulse rate $\dot{N}$ of the individual objects.  This parameter
was modeled according to:

\begin{equation}
\log\dot{N}=F \log(P)+G+\Upsilon
\end{equation}
\noindent
In the last equation $\Upsilon$ is a normal distribution with zero
mean and standard deviation 0.7 (as above, obtained from best fits). 
The fitted parameters are tabulated
in the fifth column of Table \ref{parameters}.  For each RRAT in our
model we randomly assigned a distribution for the intensity of its
individual pulses with the same three types of pulse-energy distributions used for
pulsars.  We have scaled the luminosities in the same way as
described in \S5.1.1. The spectral index of RRATs is still poorly
known, therefore we have simulated populations assuming an average
spectral index like pulsars ($\alpha=-1.82$). We note that all these
parameters are uncertain, given the small number of known sources.

\subsubsection{Monte-Carlo Simulations}

Any pulsar located in M31 must be very luminous in order to be
detected. Thus, a search for pulsars in a whole galaxy such as M31 is
an opportunity to set upper limits on the maximum luminosity of radio
pulsars in general.  To quantify this luminosity, we performed
different simulations of both pulsars and RRATs.  Each simulation had
different luminosity distributions, covering a range from that
observed in our Galaxy to a population 10 times brighter, in order to
establish which luminosity distributions would ensure a detection with
the WSRT. From now on, a model realization is defined as a single
population of pulsars. For each simulation, we produced 100
pulsar/RRAT populations. For each of these models, and based on the
luminosity distributions, we calculated the mean flux of each pulsar
given the distance to M31. We then compared these mean fluxes with the
sensitivity of our periodicity searches. Similarly we have simulated
distributions of individual pulses for each pulsar and RRAT.  The
brightest pulses of each distribution were compared with our single
pulse search sensitivity.

A detection in our models is any pulsar or RRAT that has a flux above
the minimum thresholds given in \S 3.6. Those pulsars or RRATs that
were above the minimum threshold were recorded along with their fluxes
$S$ and pulse-energy distributions. The results are summarized in
Tables \ref{models-pulsars} and \ref{models-rrats} respectively.  They
show in each column the average number of simulated pulsars detected
for each assumed distribution. In the case of single-pulse detections,
we also show the details of the power law, normal, or log-normal
distributions for those cases where we had a detection.  A general
result of our simulations is that normal radio pulsars can be more
easily detected than RRAT-like objects.

Table \ref{models-pulsars} shows that we should be able to detect at
least a few single pulses from normal radio pulsars located at the
distance of M31 only if the intensity distribution of their individual
pulses shows either a shallow power law (1$\leq \alpha \leq $2) or a
log-normal distribution.  This is the case for all known pulsars,
which have fluxes that scale proportional to $d^{-2}$. As can also be seen
in Table \ref{models-pulsars}, if the population of pulsars in M31
would be at least an average of 5 times more luminous than the one in
our Galaxy, our periodicity searches might detect a similar number of
pulsars as our single pulse searches. Thus PSR B1302$-$64 with
$L_{328}=37$ Jy kpc$^2$, the most luminous known pulsar in our Galaxy
(\citealt{mlt+78}), could be detected in M31 if it were 5 times
brighter. Based on our model of the luminosity distribution, RRATs
located in M31 are less likely to be found than pulsars, and only
those populations exhibiting more than 5 times the observed luminosity
of RRATs in our Galaxy, and having individual pulses with fluxes
distributed with either a log-normal or a shallow power law could be
detected. We note that this result was obtained by assuming that the
average spectral index of RRATs is the same as that for normal radio
pulsars. If the average index is a factor of 2 less steep we might
detect at least 1 of these objects in M31. We note that a source like
J1819$-$1458, the most luminous RRAT known with $L_{328}=5170$ mJy
kpc$^2$ (\citealt{mll+06}) would have to emit pulses about 25 times
brighter to be detected in M31. It is important to mention here that
the dwell time of the observations plays a key role in the detection
of a single pulse; the longer we observe the higher the chances of
observing a bright pulse.

The fact that we have not detected any pulsars in our periodicity searches can
be attributed to a few reasons. The first could simply be that pulsars
are not luminous enough to be detected at that distance with the
WSRT. If that is the case, these observations imply that there are no
pulsars with luminosities $L_{328}>185 \mbox{ Jy kpc}^2$.  Secondly,
the population of pulsars in M31 could be smaller than that of our
Galaxy. This argument is supported by observations made at other
wavelengths which show that M31 has a star formation rate of 0.35
M$_{\sun}$ yr$^{-1}$ (\citealt{wil03} and \citealt{bab+13}) about a
factor of 3 lower than in our Galaxy. Even though the numbers quoted
here are uncertain, this suggests that the number of neutron stars
recently produced in M31 is less than in our Galaxy. Further evidence
in favor of a smaller population of radio pulsars comes from the low
metallicity observed in M31, a factor of 2 lower than that in our
Galaxy (\citealt{scd+05} and \citealt{tdl+02}).  Photometric surveys
made in the disk of M31 have revealed a large population of old
metal-poor stars (\citealt{rkf+01}) which may mainly form black holes
when they collapse as supernov\ae ~(see \citealt{hfw+02}).  Therefore
the population of pulsars in M31 might also be older than in our
Galaxy, with most of its members being too faint to be detected.
Beaming could also cause us to miss the few pulsar candidates found by
\cite{glg}.

\subsection{Constraining the $d^{-2}$  flux law for radio pulsars}

The ``pseudo-luminosity'' of a pulsar at a given observing frequency
($L_{\nu}$) and the distance to the pulsar ($d$ in kpc) relate as:
$S_{\nu}\propto L_{\nu}/d^2$.  The mean flux density $S_{\nu}$ is
based on the integrated intensity under the pulse, averaged over the
full period.  The most widely used method for determining distance is
the ``DM-distance'', based on the NE2001 model (\citealt{cl02}) for the
distribution of free electrons in our Galaxy ($n_{\rm e}$). This model
allows distances to be estimated to an accuracy of about 20\% on
average for large samples but can give an error of 50\% or
more for individual objects. Since the model describes the entire
distribution of free electrons in our Galaxy, it is used to estimate
distances to pulsars from their measured DM.

\cite{ssm+04} made a maximum likelihood analysis using the observed
fluxes and DM-distances of pulsars. The conclusions of their analysis
are that either i) the distances are radically and consistently
incorrect by factors of 10, or ii) pulsars have a flux that falls off
more slowly with distance than expected, following a $\propto d^{-1}$
or $\propto d^{-1.5}$ power law. This discrepancy in fluxes is
explained by the presence of a component in the flux due to the
emission of superluminal polarization currents (\citealt{aas+1}).  In
order to test this claim, we performed simulations like those
described in \S5.1, but assuming that the flux scales as $S_{\nu}
\propto L_{\nu}/d^{\epsilon}$ with $\epsilon= 1$, $\epsilon= 1.5$ and
$\epsilon=2$, in order to predict the number of detectable pulsars at
the distance of M31.  Our detailed simulations paid particular
attention to the luminosity of the sources and a range of different
models for the single pulse intensity distributions.  The results are
summarized in the Tables \ref{models-pulsars} and \ref{models-rrats}
for pulsars and RRATs respectively.  Our results clearly show that if
the flux density obeys a $d^{-1}$ law, then we should have detected
many hundreds of pulsars. The fact that our observations and those
made of M33 (\citealt{mc+06}) revealed only a handful of single bright
pulses, and that our simulations using the fluxes scaled as
\cite{ssm+04} show a large number of detections, lead us to conclude
that the flux of pulsars does not scale as $d^{-1}$ or $d^{-1.5}$ but
simply as $d^{-2}$ as always assumed (unless the pulsars in M31 and
M33 are systematically far weaker than in our Galaxy). This argues
that some other effect than distance errors or a modified luminosity
law is responsible for the effect seen by \cite{ssm+04}. We suggest
that the observed deviation may be caused by an unaccounted selection
effect in the flux measurements of radio pulsar surveys.

\subsection{What is the detection at DM=54.7 pc cm$^{-3}$?}

The dispersed nature of the pulses detected at DM=54.7 pc cm$^{-3}$
and their similar properties (intrinsic width and peak flux) is
consistent with individual bright pulses observed in known pulsars.
The individual pulses show peak flux densities between 2.8--3.2 Jy and
a luminosity $>30$ Jy kpc$^{2}$ assuming they are located in our
Galaxy at the distance predicted by the NE2001 of
$d=2.9_{-2.1}^{+5.2}$ kpc. However given that the DM contribution of
our Galaxy may be as low as 45 pc cm$^{-3}$ in this direction, this
makes it possible that these pulses originate in M31, in which case
these pulses would be $\sim 70000\times$ more luminous.  Neither the
arrival-time distribution of individual pulses shown by the candidate
at DM=54.7 pc cm$^{-3}$ or the intrinsic width of the pulses is
consistent with that observed in the Crab's giant pulses, as reported
in \cite{mc+06} and \cite{ksv10}. If these pulses would be giant
pulses from a distant Crab-like pulsar, then we would have detected a
similar or higher number of pulses per hour (N$>10$ h$^{-1}$ for the
Crab) in many, if not all, of our observations.  Assuming the signals
are indeed astronomical in nature, it is thus more likely that the
source is showing normal pulsar or RRAT-like pulses in a burst, as has
been seen in other, known sources.

\subsection{Detectability of cosmological millisecond radio bursts}

Though we did not obviously detect any such events, the long dwell times of our
observations provided the potential to also detect extragalactic
millisecond radio bursts occurring at cosmological distances and in
the general direction of M31 (cf. \citealt{lbm+07}).  Here we estimate
the number of such bursts that could be observable in our
data. \cite{lbm+07} estimated the rate of such events by using the
properties of the Parkes Multibeam Survey over an area of 5 square
degrees (1/8250 of the entire sky) at any given time over a 20-day
period.  Assuming the bursts to be distributed isotropically over the
sky, they infer a nominal rate of $8250/20 \approx 225$ similar events
per day.  Based on those estimates, the expected number of bright
single bursts per day per square degree, $n_{\rm B}$, is $\sim 0.01$ bursts
day$^{-1}$ sqr deg$^{-1}$.

We have observed for a total of 80 h (3.33 days) with a
FoV $\theta = 2.57^{\circ}$ across. The area of the sky
covered by the primary beam is thus $\Omega= k [1 - \cos (\theta/2)]$
= 5.15 sqr. deg., with $k=2 \cdot 3283 \cdot \pi$. Therefore, we
estimate the total expected number of pulses to be:

\begin{equation}
n_{\rm B-M31}= \frac {0.01}{\mbox{day} \cdot \mbox{sqr. deg.}} \times \mbox{ 3.33 day} \times 5.15 \textrm{ sqr. deg.} = 0.17. 
\end{equation}
\noindent
This implies that we would likely need to observe longer to
detect such an extragalactic millisecond radio burst, assuming that a
cosmological radio burst can still be detected at 328 MHz after crossing near
M31.  The bright single pulses listed in Table
\ref{results_single_pulse1} are arguably not like the
millisecond radio burst detected by \cite{lbm+07}, because i) they
are not nearly as bright as the one reported there, ii) we do
not have precise localization of these sources and hence they may
  well originate in M31, and iii) they are likely not at cosmological
distances due to the DMs observed, which can be easily explained by the
galactic medium of M31.

\section*{Conclusions}

We have presented the most sensitive search for pulsars and radio
transients ever made in the direction of M31.  We present our
conclusions below.

\begin{enumerate}

\item We have detected a handful of single bright pulses with a DM
  that locates them beyond our Galaxy and plausibly within
  M31. Moreover we detected a few repeating bright pulses that
  consistently show the same pulse width and DM. Some of them have a
  DM consistent with that of a source at the distance of M31 while
  others could be located within our own Galaxy.

\item We performed simulations using a simple synthesis code to
  estimate the properties of a population of pulsars and RRATs in
  M31. From our simulations, we conclude that the only means to
  detect pulsars in M31 with the sensitivity of the WSRT in the
  {\tt 8gr8} mode is by detecting bright pulses described by a
  shallow power law or a log-normal distribution and assuming that the
  population of pulsars in M31 has similar characteristics to that of
  our Galaxy.  The detections reported here might be associated with
  pulsars or RRATs that show these characteristics.  Normal
    pulsars are more easily detected than RRATs because the individual
    pulses of RRATs are generally no brighter than normal pulsar
    pulses.

\item As a consequence of observing with the WSRT in the {\tt 8gr8}
  mode, we were not able to locate sporadic bursts to high accuracy
  within the primary beam. Further observations with shorter dwell
  times made with the GBT did not confirm these detections at the
  expected locations.

\item The fact that we do not detect any pulsars in our periodicity
  searches can be explained by the absence of extremely bright pulsars
  ($L_{328}>185 \mbox{ Jy kpc}^2$) and/or a smaller population of pulsars
  and RRATs in M31 compared with our Galaxy.

\item Our observations indicate that the flux of pulsars does not
  scale as $d^{-1}$ or $d^{-1.5}$ (as suggested by \citealt{ssm+04}),
  but simply as $d^{-2}$. Nevertheless there is an inconsistency that
  possibly can be attributed to selection effects in the pulsar
  surveys which show a bias towards the detection of pulses with low
  fluxes (Malmquist bias) or perhaps a combination of this effect with
  the distance of pulsars.
\end{enumerate}

In the future, WSRT, in combination with the APERTIF focal plane
array, will allow for similar surveys to the one presented here, but
with a larger instantaneous FoV and a higher observing frequency (1.4
GHz, which is likely better for detecting highly dispersed pulses).
Future searches of M31 with LOFAR are also promising
(\citealt{sha+11}), and the SKA is likely to make the first detections
of pulsars in the nearby galaxies visible from the Southern
Hemisphere.

\section*{Acknowledgments}

The Westerbork Synthesis Radio Telescope is operated by ASTRON, the
Netherlands Institute for Radio Astronomy, with support from NWO, the
Netherlands Foundation for Scientific Research.  The GBT is operated
by the National Radio Astronomy Observatory a facility of the National
Science Foundation operated under cooperative agreement by Associate
Universities, Inc.  We thank both the WSRT and GBT Operators for their
help in planning and executing these observations.  ERH acknowledges
funding from NWO and LKBF and currently is a DGAPA--UNAM Fellow.  JWTH
is an NWO Veni Fellow.

\bibliographystyle{../../Bibliography/aa}
\bibliography{../../Bibliography/journals_apj,../../Bibliography/modrefs,../../Bibliography/psrrefs,../../Bibliography/crossrefs,../../Bibliography/modjoeri}

\appendix
\section{Single bright pulses}

\begin{table*}
\begin{minipage}{150mm}

\caption{Single bright pulses observed in the direction of M31 with
  S/N$>7.0$. The pulses shown in this table were only seen once at
  this spatial location and DM. We indicate here the number of
  sub-beams in which each pulse was detected, the peak DM of each
  detection, the distance (using the NE2001 model), S/N, approximate
  pulse width ($W$), the arrival times relative to the start of the
  observation and the start time of each observation (topocentric). The last column shows the name and the number of the
  pointing in which the detections occurred (for example PNT01 3 is
  the third pointing out of four pointings made with the name PNT01
  etc). The detections in boldface are shown in
  Figs. \ref{waterfall-plot-dm139} and \ref{pulses-high-dm} and are
  representative examples of pulses with DM $> 100$~pc
  cm$^{-3}$. \label{results_single_pulse1}}

\centering
\begin{tabular}{r r r r r r r r } 
\hline
\hline
\# Sub--beams & DM~~~~~        & Distance & S/N      & Width     &  Arrival Times  & MJD    & Pointing \\
         & (pc cm$^{-3}$)  & (kpc)    &          & (ms)     &  after start Obs. ~(s)~ & at Start Obs.    &      \\
\hline
71  & 10.1 & 0.7 & 7.2 &  1.64 & 21328.5472 & 53707.63924 & PNT01 1 \\
80  & 11.8 & 0.8 & 7.1 &  1.64 & 28212.1787 & 53711.62824 & PNT01 3 \\
129 & 14.9 & 0.9 & 8.3 & 13.11 & 27872.4655 & 53711.62824 & PNT01 3 \\
 70 & 18.1 & 1.0 & 7.0 &  6.55 &  1825.6912 & 53711.62824 & PNT01 3 \\
192 & 21.6 & 1.1 & 7.2 & 26.21 &  1288.6007 & 53708.63657 & PNT02 1 \\
127 & 27.1 & 1.3 & 7.4 &  1.64 & 18938.7212 & 53713.62292 & PNT01 4 \\
108 & 29.6 & 1.4 & 7.3 &  3.28 & 23036.7858 & 53707.63924 & PNT01 1 \\
 89 & 34.8 & 1.6 & 7.1 &  1.64 & 15914.1666 & 53709.63368 & PNT01 2 \\
 84 & 38.6 & 1.8 & 7.6 &  3.28 &  7734.8040 & 53713.62292 & PNT01 4 \\
 80 & 40.0 & 1.9 & 7.3 &  3.28 & 22331.8511 & 53711.62824 & PNT01 3 \\
 67 & 40.7 & 1.9 & 7.4 &  6.55 & 13266.5088 & 53709.63368 & PNT01 2 \\
102 & 41.4 & 2.0 & 7.5 &  6.55 &  8273.8425 & 53707.63924 & PNT01 1 \\
 94 & 45.2 & 2.2 & 7.0 &  1.64 & 24262.7588 & 53709.63368 & PNT01 2 \\
130 & 45.3 & 2.2 & 7.1 &  6.55 & 27454.9585 & 53711.62824 & PNT01 3 \\
 90 & 46.6 & 2.3 & 7.1 &  6.55 & 20364.5347 & 53711.62824 & PNT01 3 \\
105 & 49.8 & 2.5 & 7.5 &  1.64 & 18631.0985 & 53709.63368 & PNT01 2 \\
109 & 50.1 & 2.5 & 7.2 &  6.55 &  2684.4553 & 53707.63924 & PNT01 1 \\
107 & 52.6 & 2.7 & 7.1 & 13.11 &  7510.3121 & 53707.63924 & PNT01 1 \\
 71 & 53.6 & 2.8 & 7.2 &  1.64 & 15459.1223 & 53717.61192 & PNT02 4 \\
108 & 56.7 & $\geq$ 3 & 7.6 &  6.55 & 4357.8804 & 53707.63924 & PNT01 1 \\
 94 & 57.1 & $\geq$ 3 & 7.1 & 13.11 & 4136.9749 & 53711.62824 & PNT01 3 \\
 92 & 57.8 & $\geq$ 4 & 7.0 &  1.64 & 26309.9242 & 53710.63113 & PNT02 2 \\
 69 & 58.1 & $\geq$ 4 & 8.4 & 26.21 & 18512.8118 & 54922.55336 & PNT03 1 \\
 77 & 62.3 & $\geq$ 4 & 7.2 &  1.64 & 18518.6699 & 53707.63924 & PNT01 1 \\
 95 & 64.0 & $>48$ & 7.1 &  1.64 &  1141.2815 & 53712.62558 & PNT02 3 \\
 96 & 64.7 & $>48$ & 7.1 &  3.28 & 20590.9174 & 53709.63368 & PNT01 2 \\
108 & 65.8 & $>48$ & 7.0 &  3.28 &  2932.2125 & 53711.62824 & PNT01 3 \\
 94 & 67.2 & $>48$ & 7.3 &  3.28 & 16470.0635 & 53709.63368 & PNT01 2 \\
 58 & 67.6 & $>48$ & 7.0 &  1.64 &  7617.9492 & 53710.63113 & PNT02 2 \\
118 & 67.5 & $>48$ & 7.4 &  6.55 & 16216.8373 & 53717.61192 & PNT02 4 \\
 17 & 68.6 & $>48$ & 7.4 &  1.64 & 26166.0776 & 53707.63924 & PNT01 1 \\
105 & 70.3 & $>48$ & 7.7 &  1.64 & 22793.9938 & 53710.63113 & PNT02 2 \\
113 & 70.7 & $>48$ & 7.0 & 13.11 &   677.8642 & 53712.62558 & PNT02 3 \\
109 & 71.0 & $>48$ & 7.7 &  3.28 & 6286.81953 & 53712.62558 & PNT02 3 \\
 67 & 87.1 & $>48$ & 7.3 &  1.64 & 23750.1026 & 53709.63368 & PNT01 2 \\
 93 & 89.9 & $>48$ & 7.7 &  1.64 & 14814.1162 & 53707.63924 & PNT01 1 \\
121 & 93.4 & $>48$ & 7.4 & 13.11 & 11957.6860 & 53711.62824 & PNT01 3 \\
{\bf{218}} & {\bf{104.6}} & $\mathbf{>48}$  & {\bf{7.9}} & {\bf{6.55}}  & {\bf{ 3436.0778}} & {\bf{53708.63657}} & {\bf{PNT02 1}} \\
{\bf{79}}  & {\bf{139.4}} & $\mathbf{>48}$  & {\bf{7.0}} &  {\bf{1.64}} & {\bf{15239.1860}} & {\bf{53709.63368}} & {\bf{PNT01 2}} \\
 61 & 189.7 & $>48$  & 7.8 & 13.11 & 27408.3190 & 54930.30243 & PNT03 2 \\
 64 & 202.2 & $>48$  & 7.7 & 13.11 & 26488.6130 & 54930.30243 & PNT03 2 \\
 81 & 258.7 & $>48$  & 9.9 & 26.21 & 18498.2226 & 54922.55336 & PNT03 1 \\
{\bf{126}} & {\bf{264.3}} & $\mathbf{>48}$  & {\bf{7.2}} & {\bf{1.64}} & {\bf{9706.9174}} & {\bf{53712.62558}} & {\bf{PNT02 3}} \\
{\bf{108}} & {\bf{267.1}} & $\mathbf{>48}$  & {\bf{7.1}} & {\bf{3.28}} & {\bf{27741.8383}} & {\bf{53712.62558}} & {\bf{PNT02 3}} \\
\hline
\hline
\end{tabular}
\end{minipage}
\end{table*} 

\begin{figure*}
\includegraphics[height=8.5cm,width=5.5cm,angle=-90]{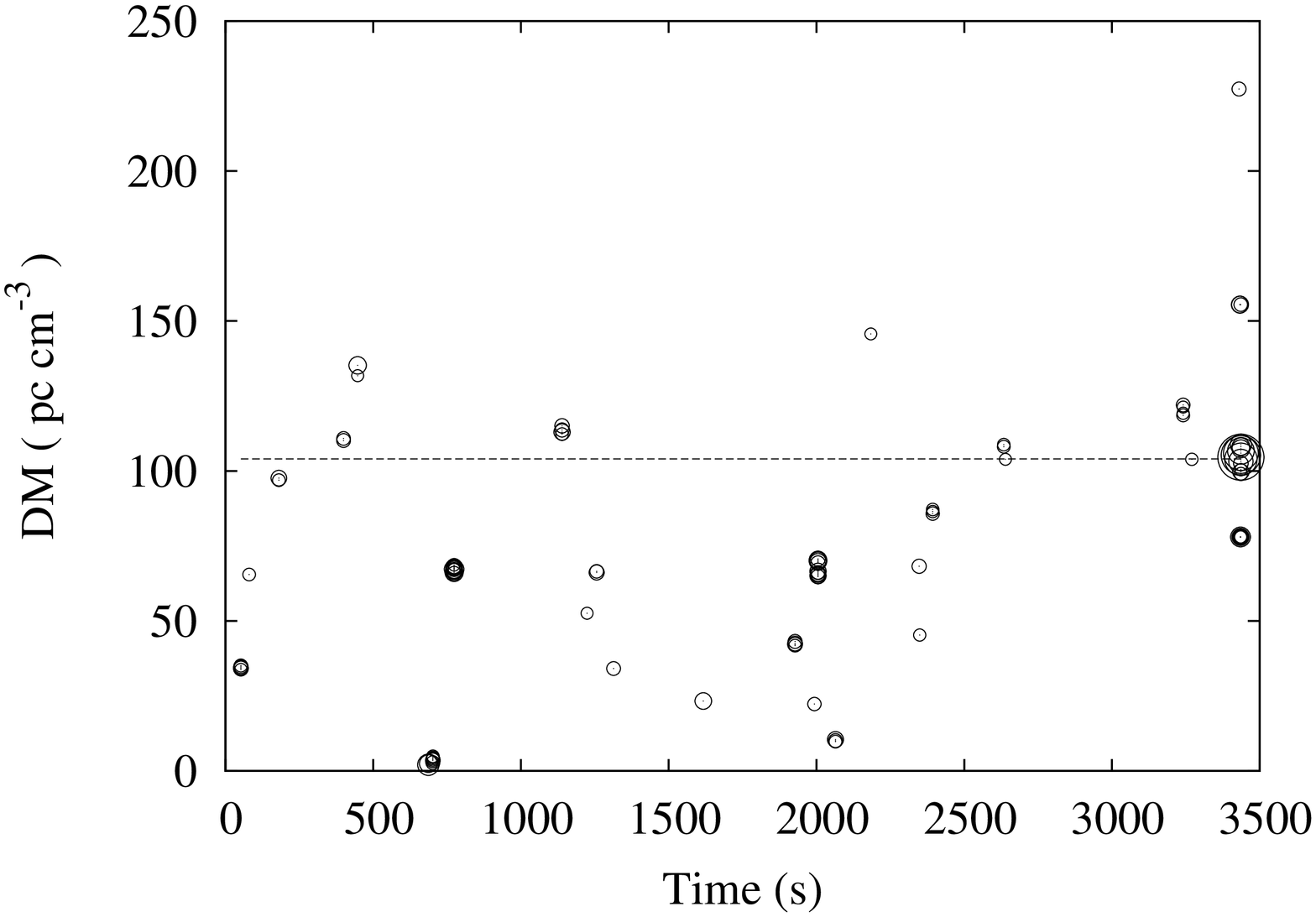}
\includegraphics[height=8.5cm,width=5.6cm,angle=-90]{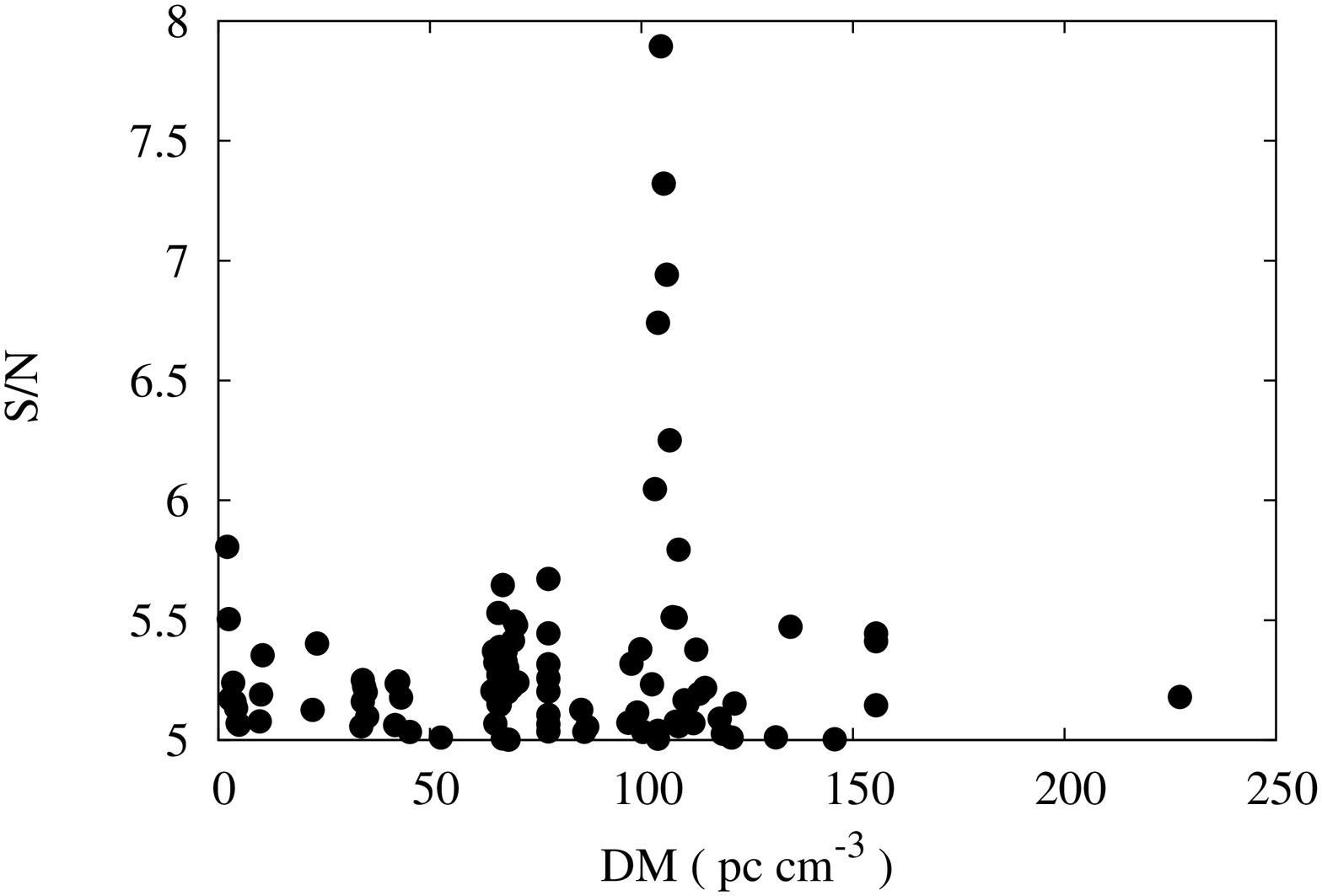}
\includegraphics[height=8.5cm,width=5.5cm,angle=-90]{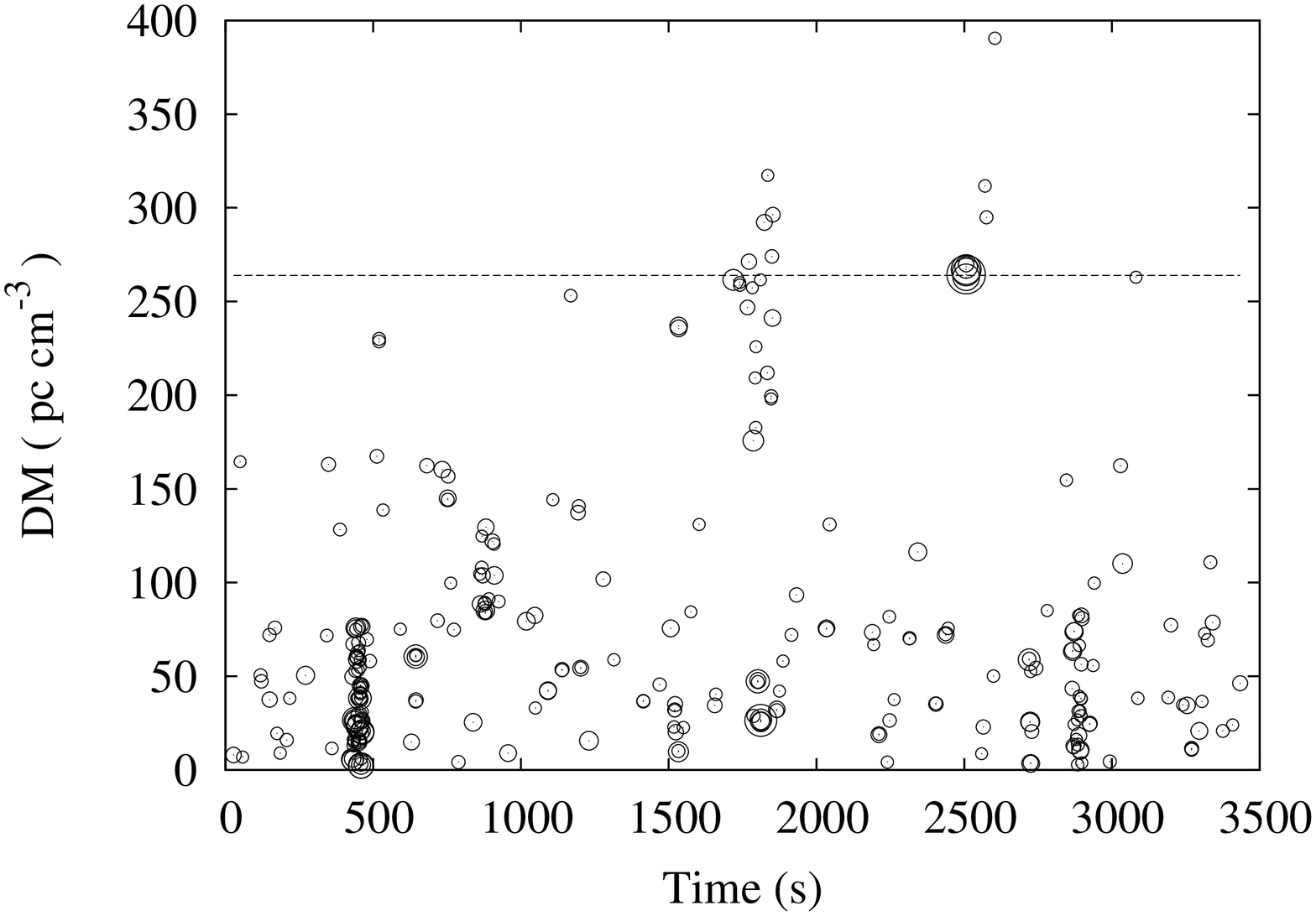}
\includegraphics[height=8.5cm,width=5.6cm,angle=-90]{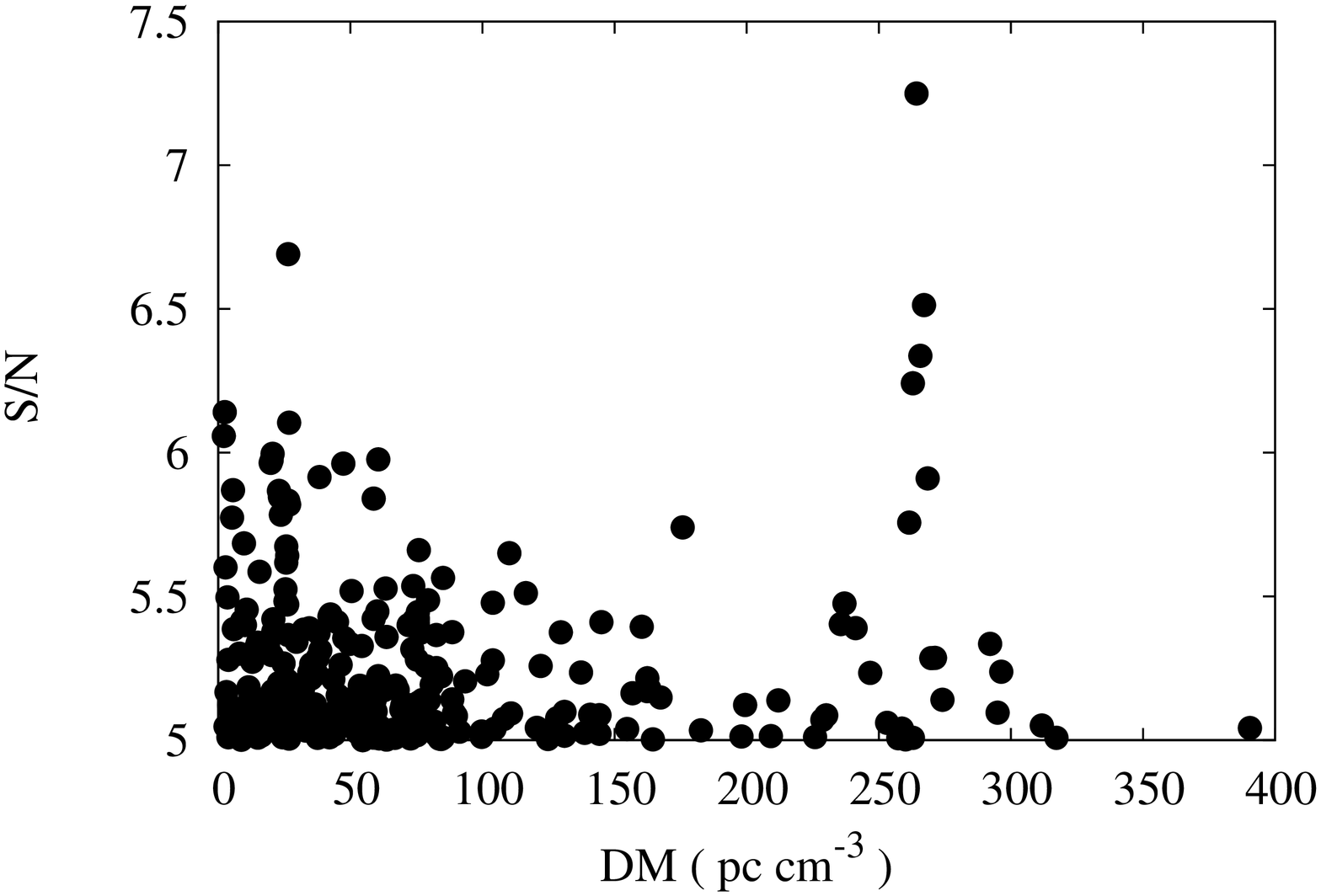}
\includegraphics[height=8.5cm,width=5.5cm,angle=-90]{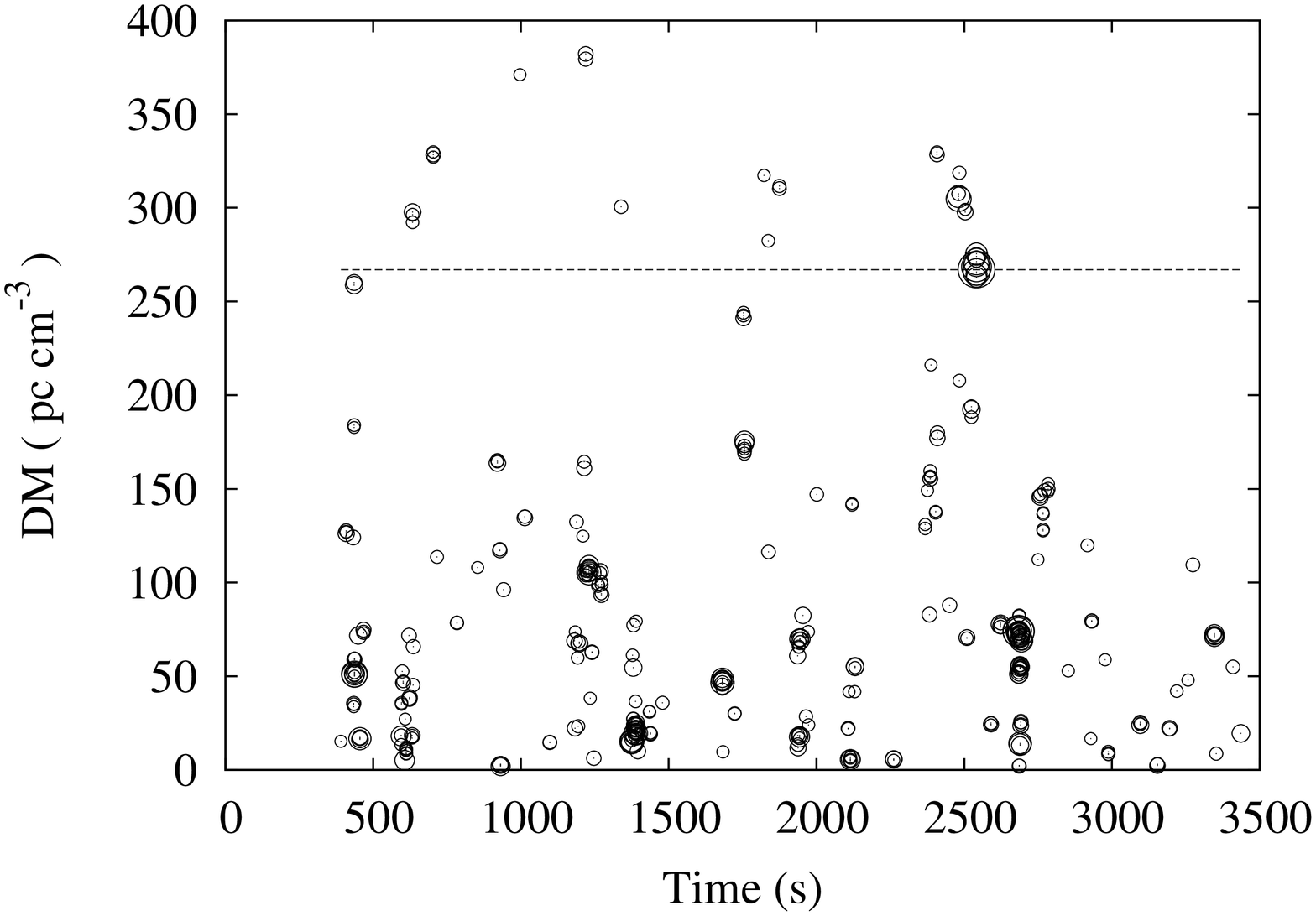}
\includegraphics[height=8.5cm,width=5.6cm,angle=-90]{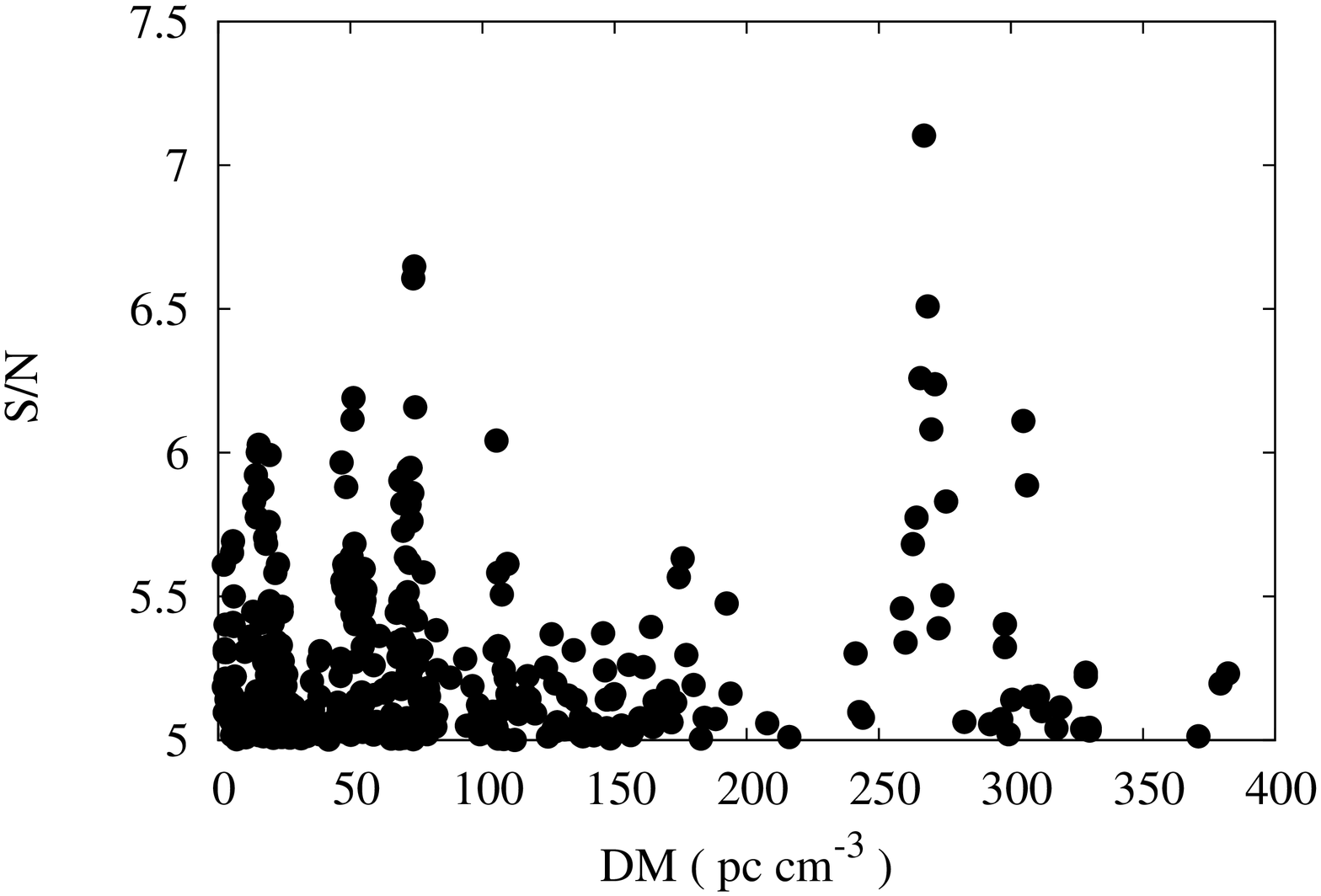}

\caption{Examples of the events detected as single bright pulses,
  which are summarized in Table~\ref{results_single_pulse1}. The
  panels show from top to bottom single bright pulses detected at DM =
  104 pc cm$^{-3}$ (t $\sim$ 3436 s) DM =264 pc cm$^{-3}$ (t $\sim$
  2506 s) and DM = 267 pc cm$^{-3}$ (t $\sim$ 2541 s).  In the
    left panels we show, for clarity, the peak DM of each detection with
    horizontal dashed lines.  The high DM of these detections suggest
that they are located outside our Galaxy. The right panels show the
S/N vs. DM plots. The $x$ axes of the right panels go only to $t=3500$
s. \label{pulses-high-dm}}
\end{figure*}

\begin{figure*}
\begin{center}
\hspace*{1.75cm}
\includegraphics[height=0.45\textwidth,width=0.40\textwidth,angle=-90]{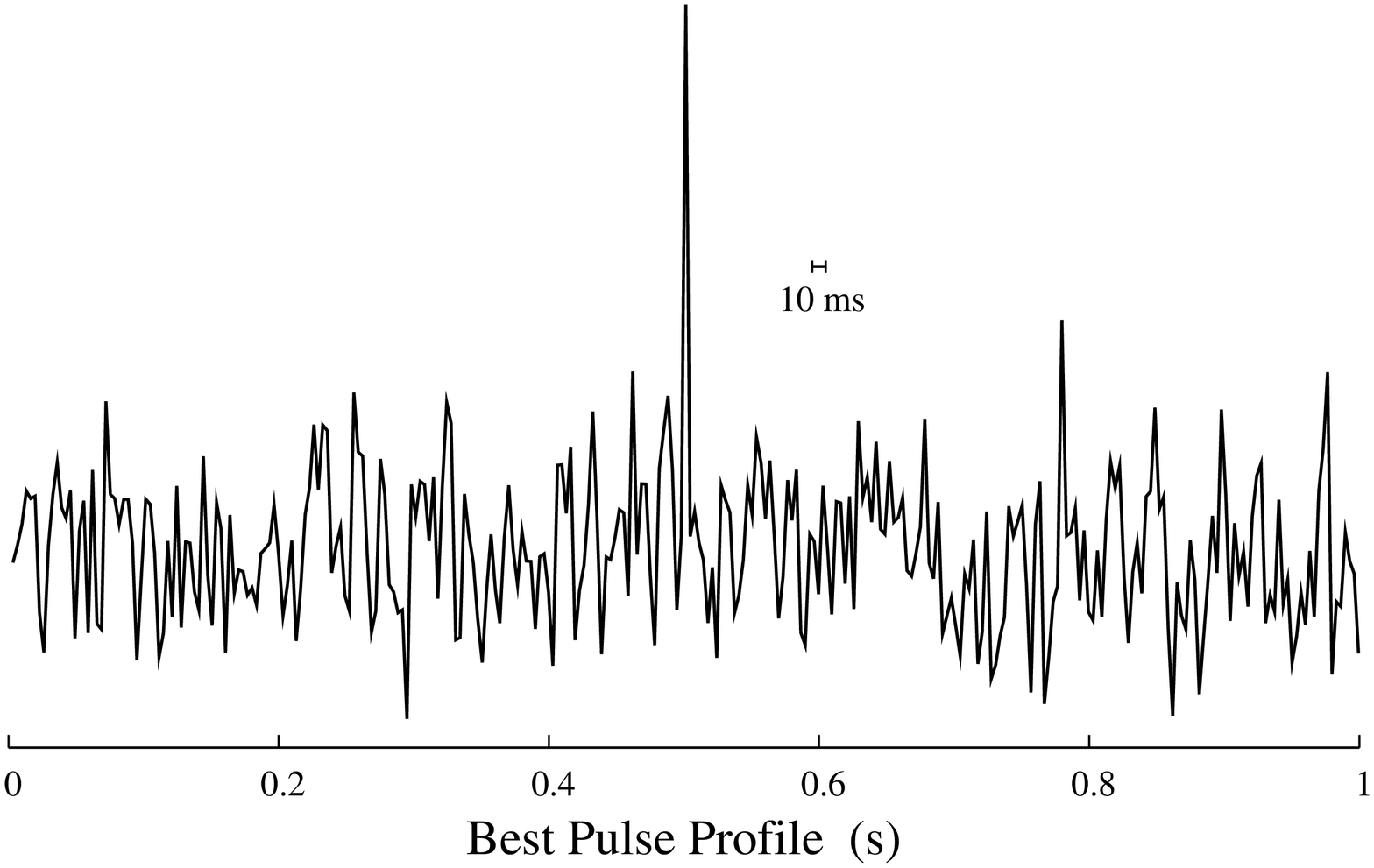}
\includegraphics[bb=166 140 484 578,height=0.55\textwidth,width=0.4\textwidth,angle=-90]{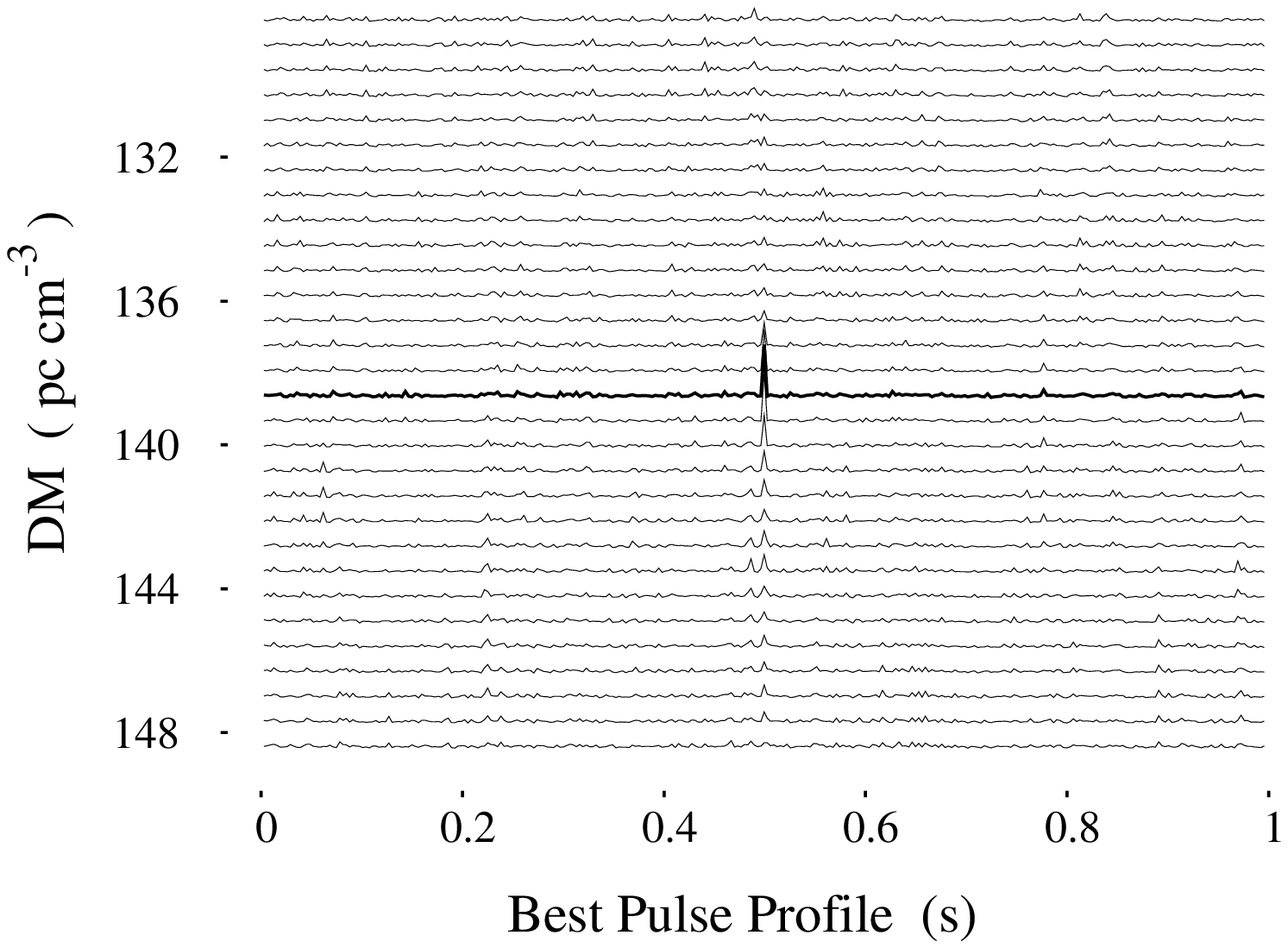}

{\caption{Pulse profile of a single pulse detected with a DM=139.41 pc
    cm$^{-3}$.  The upper panel shows the time series at the DM of the
    detection.  The lower panels show how the individual pulses
    decrease in S/N at adjacent DM trials both above and below
    DM=139.4 pc cm$^{-3}$ (thick line).  The high dispersion of this
    pulse locates it at a distance $> 48 $ kpc, therefore the pulse
    plausibly originates in M31.  The details of this detection are
    shown in Table
    \ref{results_single_pulse1}. \label{waterfall-plot-dm139}}}

\end{center}
\end{figure*}

\section{Multiple pulses at the same DM}
\begin{table*}
\begin{minipage}{150mm}

\caption{Bright pulses detected more than once at the same DM during
  80 h of observations of M31. We show the same parameters as in
  Table \ref{results_single_pulse1}, including the arrival times of
  the detections (topocentric). We stress that these
  candidates are potentially detected from the same sky direction. \label{results_single_pulse2}}

\centering
\begin{tabular}{r r r r r r r r } 
\hline
\hline
\# Beams & DM~~~~~~        & Distance & S/N      & Width      &  Arrival Times          & MJD (2000.0)  & Pointing \\
         & (pc cm$^{-3}$)  & (kpc)    &          & (ms)~~      &   after start Obs.(s)~ & at Start Obs.  &      \\
\hline
 11 & 19.9 & 1.1    & 7.7 & 3.28 & 25736.9569 & 53707.63924 & PNT01 1 \\
117 & 19.9 & 1.1    & 7.4 & 3.28 & 20047.8828 & 53713.62292 & PNT01 4 \\
\hline
108 & 63.0 & $>$ 48 & 7.1 & 1.64 &    31.3767 & 53709.63368 & PNT01 2 \\
115 & 63.0 & $>$ 48 & 7.4 & 1.64 & 15098.8447 & 53709.63368 & PNT01 2 \\
\hline
 20 & 69.3 & $>48$  & 7.0 & 1.64 & 28002.8715 & 53707.63924 & PNT01 1 \\
 59 & 69.3 & $>48$  & 7.3 & 1.64 & 12399.2815 & 53707.63924 & PNT01 1 \\
 91 & 69.3 & $>48$  & 7.2 & 1.64 & 11378.5616 & 53711.62824 & PNT01 3 \\
\hline
 91 & 72.1 & $>48$  & 7.5 & 13.11 &  566.5972 & 53709.63368 & PNT01 2 \\
 51 & 72.1 & $>48$  & 7.2 & 13.11 & 1938.6851 & 53712.62558 & PNT02 3 \\ 
\hline
{\bf{12}} & {\bf{72.4}} & $\mathbf{>48}$ & {\bf{7.4}} & {\bf{1.64}} & {\bf{27776.636314}} & {\bf{53707.63924}} & {\bf{PNT01 1}} \\
{\bf{99}} & {\bf{72.4}} & $\mathbf{>48}$ & {\bf{7.6}} & {\bf{3.28}} & {\bf{13474.783846}} & {\bf{53707.63924}} & {\bf{PNT01 1}} \\
\hline
\hline
\end{tabular}
\end{minipage}
\end{table*}

\begin{figure*}
\begin{minipage}{0.9\textwidth}
\includegraphics[height=0.5\textwidth,width=0.4\textwidth,angle=-90]{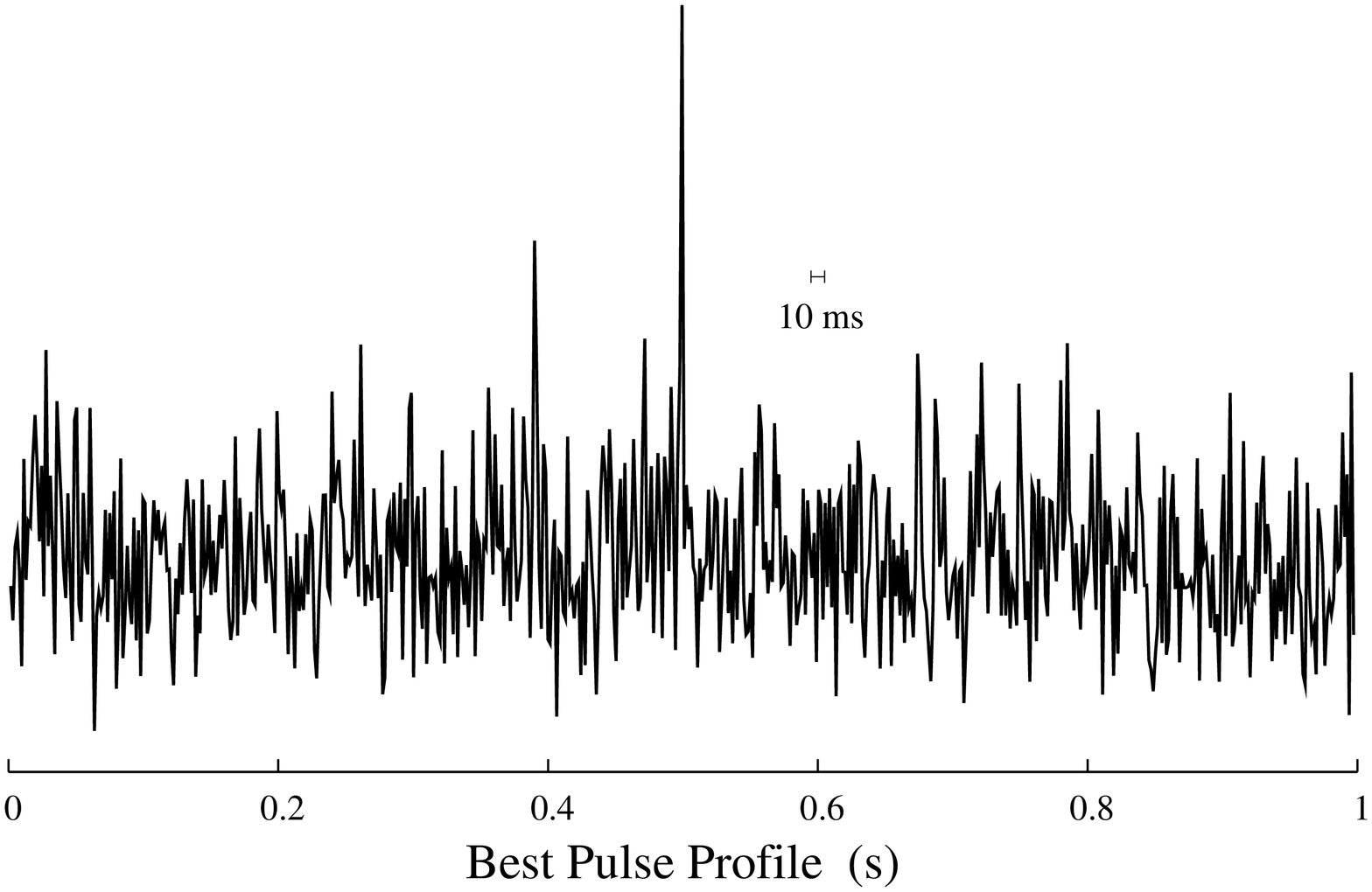}
\hspace*{0.75cm}
\includegraphics[height=0.5\textwidth,width=0.4\textwidth,angle=-90]{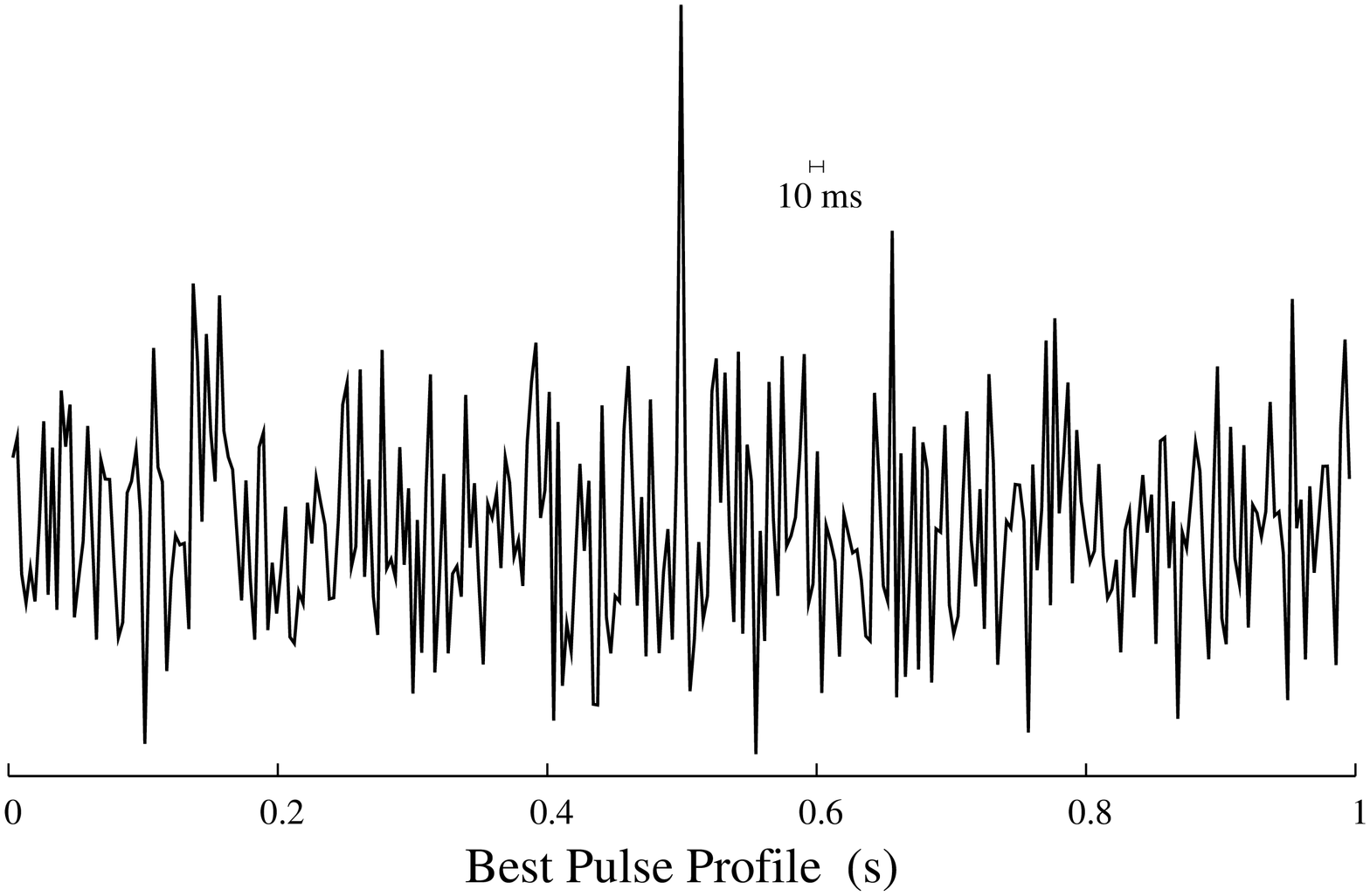}
\vspace*{1cm}
\end{minipage}

\begin{minipage}{\textwidth}
\hspace*{-1.0cm}
\includegraphics[bb=166 139 484 578,height=0.55\textwidth,width=0.4\textwidth,angle=-90]{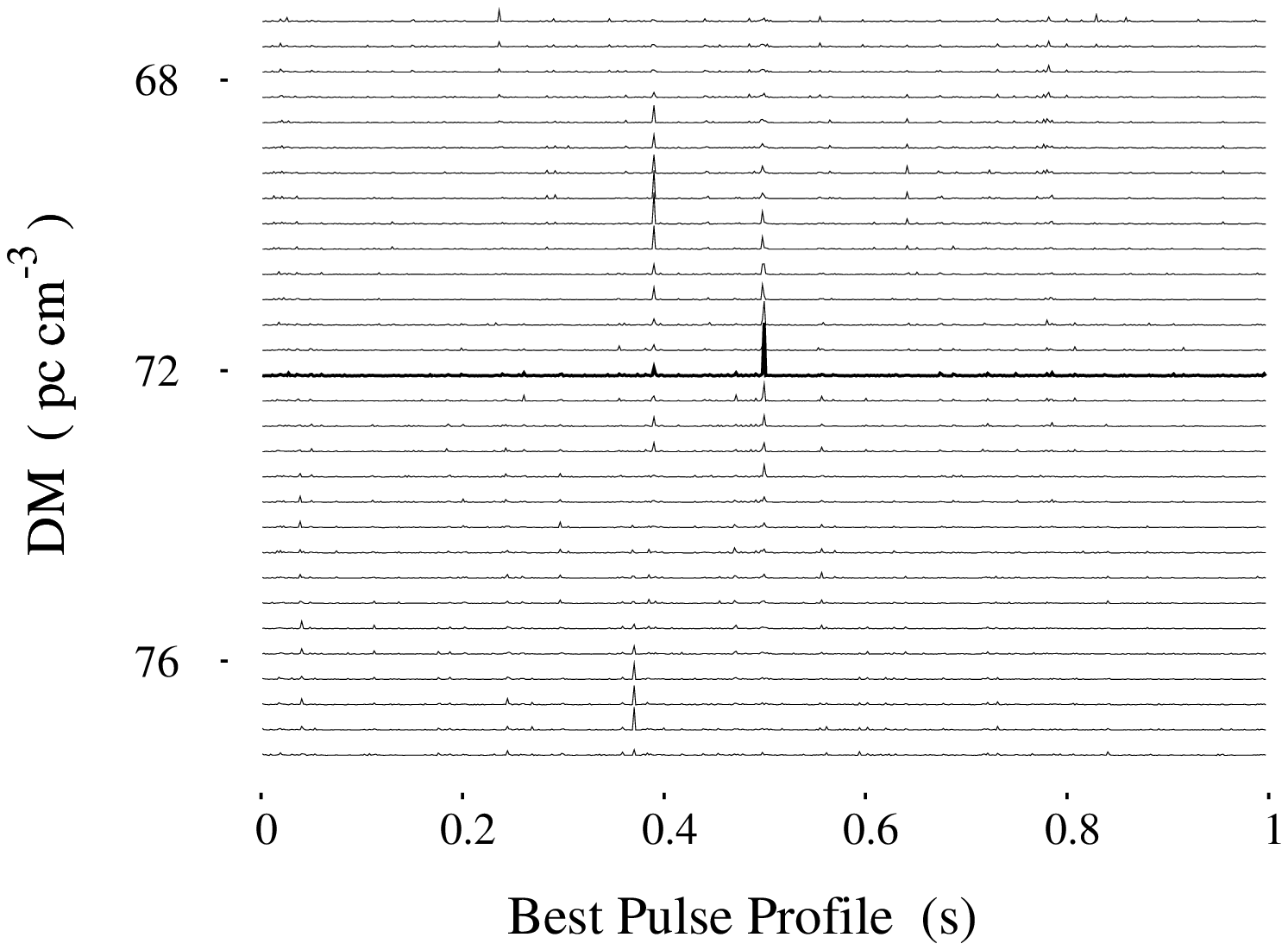}
\hspace*{-1.0cm}
\includegraphics[bb=166 140 484 578,height=0.55\textwidth,width=0.4\textwidth,angle=-90]{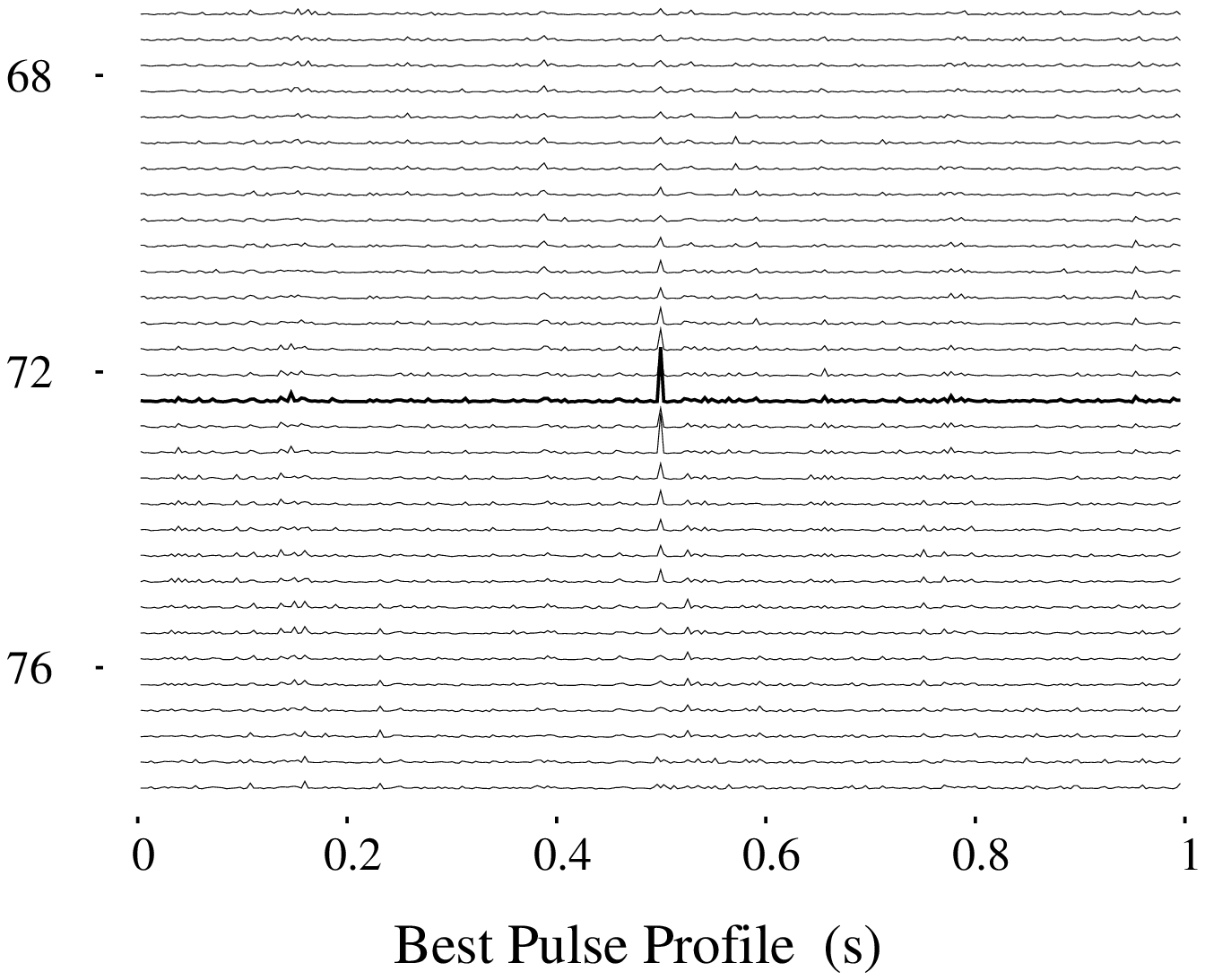}
\end{minipage}

{\caption{Pulse profile of two pulses detected at DM=72.4 pc cm$^{-3}$
    at different times, as shown in Table B1. For the two pulses, the
    upper panels show the time series at the DM of the detections. The
    lower panels show how the individual pulses decrease in S/N at
    adjacent DM trials both above and below DM=72.4 pc cm$^{-3}$
    (thick line). The high DMs of these pulses locates them at
    a distance $> 48 $ kpc; therefore these pulses plausibly
    originate in M31.  The details of this detection are shown in
    boldface in Table \ref{results_single_pulse2}.}}

\end{figure*}

\label{lastpage}
\end{document}